\documentclass[10pt,conference]{IEEEtran}

\usepackage{tikz}
\newcommand*\circled[1]{\tikz[baseline=(char.base)]{
            \node[shape=circle,draw,inner sep=1pt] (char) {#1};}}

\usepackage{cancel}
\usepackage{xcolor}
\usepackage{algorithm}
\usepackage{algorithmic}
\usepackage{bm}

\usepackage{cite}
\usepackage{amsmath,amssymb,amsfonts}
\usepackage{algorithmic}
\usepackage{graphicx}
\usepackage{textcomp}
\usepackage{xcolor}
\usepackage[hyphens]{url}
\usepackage{fancyhdr}
\usepackage{hyperref}

\usepackage{nicefrac}

\usepackage[capitalize]{cleveref} 

\usepackage{xspace}
\usepackage{enumitem}

\usepackage{tcolorbox}

\usepackage{multicol}
\usepackage{multirow}
\usepackage{makecell}
\usepackage{booktabs}

\usepackage{afterpage}

\usepackage{siunitx}

\definecolor{olivegreen}{rgb}{0, 0.6, 0}
\definecolor{redorange}{HTML}{FF5349}
\definecolor{blue(ncs)}{rgb}{0.0, 0.53, 0.74}
\definecolor{navy}{HTML}{273BE2}

\definecolor{black}{HTML}{000000}
\definecolor{white}{HTML}{ffffff}
\definecolor{color1}{HTML}{ACE5EE}
\definecolor{color2}{HTML}{0093AF}
\definecolor{color3}{HTML}{CC0000}
\definecolor{color4}{HTML}{0087BD}
\definecolor{color5}{HTML}{333399}
\definecolor{color6}{HTML}{20B2AA}

\usepackage{pifont}

\newcommand{\thiswork}{\textsc{LoCaLUT}\xspace}

\newcommand{\Sort}{LUT Canonicalization\xspace}
\newcommand{\sort}{LUT canonicalization\xspace}

\newcommand{\lut}{canonical LUT\xspace}

\newcommand{\Reorder}{Reordering LUT\xspace}
\newcommand{\reorder}{reordering LUT\xspace}

\newcommand{\Dataflow}{LUT Slice Streaming\xspace}
\newcommand{\dataflow}{LUT slice streaming\xspace}

\newcommand{\JL}[1]{{\color{cyan}[\textbf{\sc JLee}: \textit{#1}]}}
\newcommand{\JH}[1]{{\color{orange}[\textbf{\sc JHong}: \textit{#1}]}}
\newcommand{\jh}[1]{{\color{orange}[\textbf{\sc JHong}: \textit{#1}]}}
\newcommand{\CS}[1]{{\color{blue(ncs)}[\textbf{\sc CShin}: \textit{#1}]}}

\newcommand{\SJ}[1]{{\color{blue}[\textbf{\sc SKim}: \textit{#1}]}}
\newcommand{\SN}[1]{{\color{olive}[\textbf{\sc SNoh}: \textit{#1}]}}
\newcommand{\THK}[1]{{\color{magenta}[\textbf{\sc TKwon}: \textit{#1}]}}

\iffalse 
\renewcommand{\JL}[1]{}
\renewcommand{\JH}[1]{}
\renewcommand{\jh}[1]{}
\renewcommand{\CS}[1]{}
\renewcommand{\SJ}[1]{}
\renewcommand{\SN}[1]{}
\renewcommand{\THK}[1]{}
\else
\fi

\definecolor{olivegreen}{rgb}{0, 0.6, 0}
\definecolor{redorange}{HTML}{FF5349}
\definecolor{blue(ncs)}{rgb}{0.0, 0.53, 0.74}
\definecolor{navy}{HTML}{273BE2}
\definecolor{greenyellow}{HTML}{DDE576}
\usepackage{xcolor}

\usepackage{hyperref}
\usepackage{marginnote}

\newcommand{\hpcayear}{2026}

\newcommand{\hpcasubmissionnumber}{1909}
\title{\thiswork: Harnessing Capacity–Computation Tradeoffs for LUT-Based Inference in DRAM-PIM}

\def\hpcacameraready{} 

\newcommand\hpcaauthors{
Junguk Hong\IEEEauthorrefmark{2},
Changmin Shin\IEEEauthorrefmark{2},
Sukjin Kim\IEEEauthorrefmark{2},
Si Ung Noh\IEEEauthorrefmark{2},
Taehee Kwon\IEEEauthorrefmark{2},
Seongyeon Park\IEEEauthorrefmark{2}, \\
Hanjun Kim\IEEEauthorrefmark{3},
Youngsok Kim\IEEEauthorrefmark{4},
and
Jinho Lee\IEEEauthorrefmark{2}
}
\newcommand\hpcaaffiliation{
\IEEEauthorrefmark{2}\textit{Department of Electrical and Computer Engineering, Seoul National University} \\
\IEEEauthorrefmark{3}\textit{School of Electrical and Electronic Engineering, Yonsei Yniversity} \\ 
\IEEEauthorrefmark{4}\textit{Department of Computer Science and Engineering, Yonsei University}

}
\newcommand\hpcaemail{$\{$junguk16, scm8432, iamksj1212, siung98, jessica314, syeonp$\}$@snu.ac.kr, \\ $\{$hanjun, youngsok$\}$@yonsei.ac.kr, 
leejinho@snu.ac.kr
}

\def\aeopen{}           

\author{
  \ifdefined\hpcacameraready
    \IEEEauthorblockN{\hpcaauthors{}}
      \IEEEauthorblockA{
        \hpcaaffiliation{} \\
        \hpcaemail{}
      }
  \else
    \IEEEauthorblockN{\normalsize{HPCA \hpcayear{} Submission
      \textbf{\#\hpcasubmissionnumber{}}} \\
      \IEEEauthorblockA{
        Confidential Draft \\
        Do NOT Distribute!!
      }
    }
  \fi 
}

\fancypagestyle{camerareadyfirstpage}{%
  \fancyhead{}
  
  \fancyhead[C]{
    \ifdefined\aeopen
    \parbox[][12mm][t]{13.5cm}{\hpcayear{} IEEE International Symposium on High-Performance Computer Architecture (HPCA)}    
    \else
      \ifdefined\aereviewed
      \parbox[][12mm][t]{13.5cm}{\hpcayear{} IEEE International Symposium on High-Performance Computer Architecture (HPCA)}
      \else
      \ifdefined\aereproduced
      \parbox[][12mm][t]{13.5cm}{\hpcayear{} IEEE International Symposium on High-Performance Computer Architecture (HPCA)}
      \else
      \parbox[][0mm][t]{13.5cm}{\hpcayear{} IEEE International Symposium on High-Performance Computer Architecture (HPCA)}
    \fi 
    \fi 
    \fi 
    \ifdefined\aeopen 
      \includegraphics[width=12mm,height=12mm]{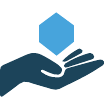}
    \fi 
    \ifdefined\aereviewed
      \includegraphics[width=12mm,height=12mm]{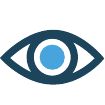}
    \fi 
    \ifdefined\aereproduced
      \includegraphics[width=12mm,height=12mm]{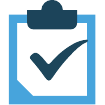}
    \fi
  }
  \fancyfoot[C]{}
}
\begin{document}

\pagenumbering{empty}

\maketitle

\ifdefined\hpcacameraready 
  \thispagestyle{camerareadyfirstpage}
  \pagestyle{empty}
\else
  \thispagestyle{plain}
  \pagestyle{plain}
\fi

\newcommand{\hpcaheight}{0mm}
\ifdefined\eaopen
\renewcommand{\hpcaheight}{12mm}
\fi
  
\pagenumbering{arabic}

\setlength{\headheight}{49pt}


\begin{abstract}

Lookup tables (LUTs) have recently gained attention as an alternative compute mechanism that maps input operands to precomputed results, eliminating the need for arithmetic logic. LUTs not only reduce logic complexity, but also naturally support diverse numerical precisions without requiring separate circuits for each bitwidth---an increasingly important feature in quantized DNNs. This creates a favorable tradeoff in PIM: memory capacity can be used in place of logic to increase computational throughput, aligning well with DRAM-PIM architectures that offer high bandwidth and easily available memory but limited logic density.

In this work, we explore this capacity-computation tradeoff in LUT-based PIM designs, where memory capacity is traded for performance by packing multiple MAC operations into a single LUT lookup.
Building on this insight, we propose \thiswork, a PIM-based design for efficient low-bit quantized DNN inference using operation-packed LUTs. 
First, we observe that these LUTs contain extensive redundancy and introduce \sort, which eliminates duplicate entries to reduce LUT size. 
Second, we propose \reorder, a lightweight auxiliary LUT that remaps weight vectors to their canonical form required by \sort with a simple LUT lookup.
Third, we propose \dataflow, a novel execution strategy that exploits the DRAM-buffer hierarchy by streaming only relevant LUT columns into the buffer and reusing them across multiple weight vectors.
{Evaluated on a real system based on UPMEM devices, we demonstrate a geometric mean speedup of 1.82$\times$ across various numeric precisions and DNN models.}
We believe \thiswork opens a path toward scalable, low-logic PIM designs tailored for LUT-based DNN inference.
Our implementation of \thiswork is available at \url{https://github.com/AIS-SNU/LoCaLUT}.

\end{abstract}


\section{Introduction}\label{sec:intro}

In recent decades, numerous processing-in-memory (PIM) architectures~\cite{truepim, hbmpim, gddraim, attacc, neupims, ianus, piccolo, gradpim, trim, psyncpim, samsung-pim, fala, hbmpim_isscc, lazypim, aespa, pidcomm} have demonstrated outstanding performance in mitigating the memory wall problem.  
By leveraging abundant internal memory bandwidth, PIM architectures can deliver significantly higher performance for data-intensive applications.  
Consequently, numerous PIM designs have been proposed for highly memory-bound workloads such as vector operations~\cite{gradpim}, matrix-vector multiplications (GEMV)~\cite{hbmpim, gddraim}, reductions~\cite{trim}, and sparse matrix algebra~\cite{psyncpim, piccolo}, which serve as fundamental building blocks for many workloads, especially for DNN inference.


However, applying PIM to more compute-intensive components of DNN inference, such as matrix-matrix multiplications (GEMM), remains challenging due to the high cost of integrating arithmetic units.
Unlike traditional processor cores fabricated in logic-optimized nodes, DRAM dies are manufactured with capacity-oriented processes, making logic circuits area- and energy-inefficient. 
As a result, many commercial PIM systems~\cite{truepim, hbmpim, aim} integrate only a small number of fixed-precision arithmetic units within the constrained area available on the DRAM die.
This not only limits arithmetic throughput, but also poses challenges for supporting modern quantized DNNs, which rely on a variety of low-bit formats \cite{lowbitquant, 2bitquant, 2bitquant_2, 3bitquant, 4bitquant, sub4_1, sub4_2, mx4, nxfp, amxfp4}.

To address this issue, lookup table (LUT)-based matrix multiplication for DNN inference~\cite{lutnn, lutdla, pimdl, luttensorcore, tmac} could be a promising alternative. 
By precomputing and storing these results offline, LUTs eliminate the need for complex multipliers, enabling fast and area-efficient implementations.
This paradigm is especially attractive for low-bit DNN inference, where the number of unique multiplication outcomes is limited. 
Moreover, it aligns naturally with PIM architectures by repurposing their memory capacity for computation.

In this work, we aim to push the idea further by allowing a tradeoff between capacity and computation.
Inspired by recent LUT designs~\cite{luttensorcore, tmac}, we employ a design where multiple operations are packed into a single LUT lookup. 
While this increases the size of the LUT, DRAM has a relatively large capacity with high internal bandwidth, making this approach appealing for DRAM-PIM architectures.
Furthermore, since the shape and contents of LUTs can be reconfigured, this provides a flexible design for quantized DNN inference.

Building on this strategy, we propose \thiswork, a design that capitalizes on this tradeoff to deliver high-throughput DNN inference in DRAM-PIM. 
\thiswork introduces a suite of additional optimization techniques that enable dense, reusable, and capacity-efficient LUT execution.


First,  we introduce \sort, a compact LUT design that eliminates redundant entries to reduce LUT capacity requirements, enabling more multiplications to be packed into a single lookup.
We observe that operation-packed LUTs inherently suffer from redundancy due to the permutation of the packed input values.
To address this, \sort only stores unique LUT entries by canonicalizing the input indices, effectively removing duplicates.

Second, to minimize the runtime overhead of computing sorted indices, we further propose \reorder, an additional lightweight auxiliary LUT that transforms the indices into a canonical format in a single LUT lookup.
Together, these techniques substantially reduce LUT size while enabling efficient execution of multiple operation-packed LUTs, thereby improving overall compute throughput.

Third, we propose \dataflow, a novel execution strategy that leverages the DRAM-buffer hierarchy.
\dataflow uses a LUT sized to fit a DRAM bank, and streams only relevant LUT columns into the buffer.
By co-designing the dataflow together to use them across multiple weight vectors, this maximizes the performance benefits. 

To demonstrate the effectiveness of our design, we implement \thiswork on a system that includes 16 UPMEM DIMMs \cite{truepim}.
The results show that \thiswork delivers performance improvements of up to 1.82$\times$ for low-bit quantized DNN models, 
compared to the prior LUT approaches.


Our contributions can be summarized as follows:
\begin{itemize}
    \item We identify a capacity-computation tradeoff unique to LUT-based inference on DRAM-PIM and design \thiswork, a memory-centric execution strategy to exploit it.
    \item We introduce three key techniques to drive efficient tradeoff for \thiswork: \sort, \reorder, and \dataflow that together reduce LUT size, avoid runtime sorting, and improve reuse.
    \item We implement \thiswork on a real PIM system and demonstrate 1.82$\times$ speedup over baseline LUT designs.
\end{itemize}

\section{Background}
\begin{figure}
    \centering
    \includegraphics[width=\columnwidth]{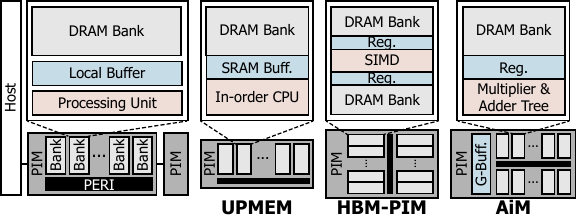}
    \caption{
    DRAM-PIM architecture abstraction. Conventionally, bank-level PIM architectures involve local buffers and processing units near each DRAM bank. 
    }
    \label{fig:pim_arch}
\end{figure}

\subsection{DRAM-PIM Architecture}
\label{sec:bg:pim}
The DRAM-based PIM (DRAM-PIM) architecture has been explored for several decades~\cite{execube,iram, flexram, smartmem, yukon, pnmpudpim}. 
While there are many variants of DRAM-PIM (e.g., processing using memory~\cite{rowclone, ambit, simdram, mimdram}, processing near memory~\cite{chameleon, tensordimm, axdimm, trim, asyncdimm}), we specifically focus on near-bank PIM~\cite{bufcmp, hbmpim, gradpim, aim, truepim}, which attaches processing units to directly feed data from DRAM banks. 

\cref{fig:pim_arch} shows a general architecture of such near-bank DRAM-PIM. 
Each DRAM channel consists of multiple banks that operate independently while sharing the off-chip I/O.
On each DRAM bank, a processing unit is attached. 
Because the bandwidth of a single bank is often comparable to that of an entire chip, this can provide a significant bandwidth advantage.

The details of the processing engine vary per design, but they typically contain a processing unit and a small local buffer near each DRAM bank. 
{In UPMEM~\cite{truepim}, the processing unit is an in-order CPU, each of which is connected to a 64KB SRAM-based local buffer.}
In HBM-PIM~\cite{hbmpim} and AiM~\cite{aim}, the processing units are parallel MAC engines, where the local buffers are per-unit registers with an additional SRAM global buffer in AiM architecture. 
In some architectures, the set of sense amplifiers is regarded as the local buffer~\cite{mvid, bufcmp}. 

The DRAM technology node is fundamentally ill-suited for logic circuit implementation due to the slower transistors, fewer metal layers, and lower logic density.
Thus, primarily due to the area constraints~\cite{hbmpim, hbmpim_isscc}, they typically support only a limited set of numeric precisions, especially for multiplications. 
In the case of UPMEM, only 8-bit integer multiplications are natively supported, where HBM-PIM and AiM only support fp16 and bf16, respectively.
Such a limitation could be potentially handled with LUT-based designs, because LUT contents can be rewritten according to different formats.

\subsection{LUT-based Matrix Multiplication}
\label{sec:bg:lut}

LUT-based computation has emerged as an attractive approach for accelerating low-bit DNN inference, especially for costly dot products~\cite{pimdl, luttensorcore, tmac}.
Rather than performing multiplications and additions using arithmetic circuits within the processing units, this method precomputes dot products and stores them in lookup tables (LUTs).
During inference, the matrix multiplication is converted to a series of LUT accesses followed by simple additions of the retrieved results.
This approach shifts the computational bottleneck from arithmetic operations to memory access, effectively transforming a compute-intensive workload into a memory-intensive one.

A few recent designs, LUT Tensor Core~\cite{luttensorcore} and T-MAC~\cite{tmac}, adopt bit-serial designs that restrict weight precision to 1-bit, significantly reducing the LUT input space.
However, these approaches require generating LUTs for each activation at runtime, incurring non-negligible performance overhead. 
Alternatively, PIM-DL~\cite{pimdl} and LUT-DLA~\cite{lutdla} leverage product quantization to reduce arithmetic complexity by approximating dot products through centroid lookups.
Nevertheless, this requires additional processing on the host to determine the best-matching centroid, introducing a new computational bottleneck.





\begin{figure}
    \centering
    \includegraphics[width=\columnwidth]{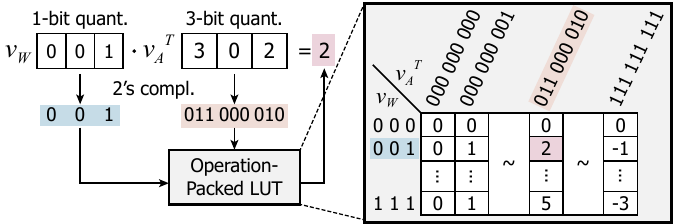}
    \caption{
    Operation-packed LUT with packing degree $p=3$.
    }
    \label{fig:packed_lut}
\end{figure}


%
%

\section{Capacity-Computation tradeoff for LUTs}
\label{sec:prelim}

\subsection{Operation-Packed LUTs}
\label{sec:prelim:oppack}
While existing LUT designs often trade one LUT lookup for a single operation (e.g., multiplication)~\cite{lt-pim} or split an operation into multiple lookups~\cite{luttensorcore, tmac}, this is not the only viable direction for the tradeoff. 
For example, to reduce the LUT size, \cite{luttensorcore, tmac} serializes weights into bits and merges several of those into a single lookup. The resulting outputs are then left-shifted and accumulated toward the final result.

{
In contrast, DRAM-PIM provides more memory capacity compared to existing accelerators that employ operation-packed LUTs.
This increased capacity enables the adoption of an \emph{operation-packed LUT} strategy specifically tailored for inner product operations.
}
In this approach, multiple operations are packed within a single LUT lookup, increasing the LUT size in exchange for higher computational throughput.
As shown in \cref{fig:packed_lut}, the inner product of multiple weights (e.g., 0, 0, 1) and activations (e.g., 3, 0, 2) is encoded into a single LUT entry, leveraging the fact that the three multiplications are accumulated to form a single output value.
The result, computed as $0 \times 3 + 0 \times 0 + 1 \times 2 = 2$, is stored in the LUT and can be retrieved using a packed index derived from the weight vector [001] and the activation vector [011 000 010].

This strategy allows trading memory capacity for computational throughput.
Specifically, computing the inner product of a $b_w$-bit weight value and a $b_a$-bit activation value for $b_o$ bytes output requires an LUT with $b_o\cdot2^{b_w + b_a}$ bytes.
By extending this to $b_o\cdot2^{(b_w + b_a) \cdot p}$ bytes, we perform $p$ (packing degree) such operations in a single LUT lookup.
{
While this technique 
introduces exponential growth in LUT size--making it impractical for on-chip accelerators~\cite{lutdla, luttensorcore}--it presents an appealing opportunity for DRAM-based PIM architectures.
}
Since the capacity growth is exponential, it becomes crucial to reduce its size as much as possible, and determine where to store the LUT (i.e., between the DRAM bank and the local buffer).
%
%
In the remainder of this section, we explore a few trivial design candidates for applying operation-packed LUT on DRAM-PIM.




\begin{figure}
    \centering
    \includegraphics[width=\columnwidth]{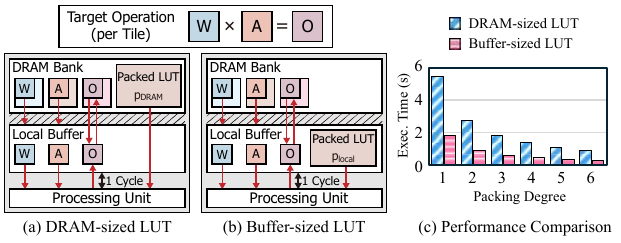}
    \caption{Baseline LUT-based approaches to perform matrix multiplications in DRAM-PIM. (a) Placing LUTs in DRAM. (b) Placing LUTs in the SRAM buffer. (c) Comparison results of the baseline approaches.}
    \label{fig:lut_baselines}
\end{figure}

\subsection{Candidate Designs}
\label{sec:prelim:candidates}
\cref{fig:lut_baselines}(a) and (b) illustrate two candidate approaches for performing matrix multiplications. 
In a DRAM-PIM, we can utilize two entities for storing the LUT: the DRAM bank and the local buffer. 
We explore two designs regarding where to store the LUT: the DRAM bank or the local buffer.
For simplicity, we assume that the inputs and outputs of a tile fit the local buffer. 

\subsubsection{DRAM Bank-sized LUT}
\label{sec:prelim:candidates:dram}
A straightforward approach is to fit the LUT in the DRAM bank, as illustrated in \cref{fig:lut_baselines}(a). 
{
Assuming a 64 MB bank size of UPMEM, we can fit an LUT with a relatively large packing degree $p_{{}_{DRAM}}$ between 1 and 6, when $b_{w}=1$ and $b_{a}=3$.
}
However, this design suffers from a high per-lookup cost due to the large DRAM bank access time compared to that of the local buffer.

\subsubsection{Buffer-sized LUT}
\label{sec:prelim:candidates:buffer}
To address the large LUT access cost of the first design, an alternative would be to use an LUT that fits the remaining space of the local buffer, as in \cref{fig:lut_baselines}(b).
Assuming around half of the 64 KB capacity of UPMEM can be used for LUTs, we obtain a packing degree $p_{{}_{local}}$ between 1 and 3. Although this design can fully utilize the high SRAM bandwidth, we cannot fully exploit the advantages of using PIM, such as its high internal bandwidth.  



\subsection{Comparison of the Candidates.}
\cref{fig:lut_baselines}(c) presents a comparison of LUT placement designs using a small-scale GEMM experiment.
The packing degree varies from 1 to 6, using two matrices of size 512$\times$512 with $b_{w}=1$, and $b_{a}=3$.
The results indicate that the local-buffer LUT consistently outperforms the DRAM-based LUT across all packing degrees. 
This is because, unlike DRAM banks, which are limited by lower bandwidth and the amortized row activation overhead, the local buffer is implemented using SRAMs or registers, enabling a single-cycle access with minimal latency.

Note that this is not an entirely fair comparison, as the DRAM bank-sized LUT does not occupy the local buffer, leaving room for a larger matrix tile to be placed in the buffer.
Nonetheless, further experiments and corresponding performance models (see \cref{sec:design:model}) demonstrate that the local buffer-sized LUT outperforms the alternative design, and thus we choose the buffer-sized LUT as the base design of \thiswork to start with.

\section{\thiswork Optimizations}
\label{sec:design}

\begin{figure}
    \centering
    \includegraphics[width=\columnwidth]{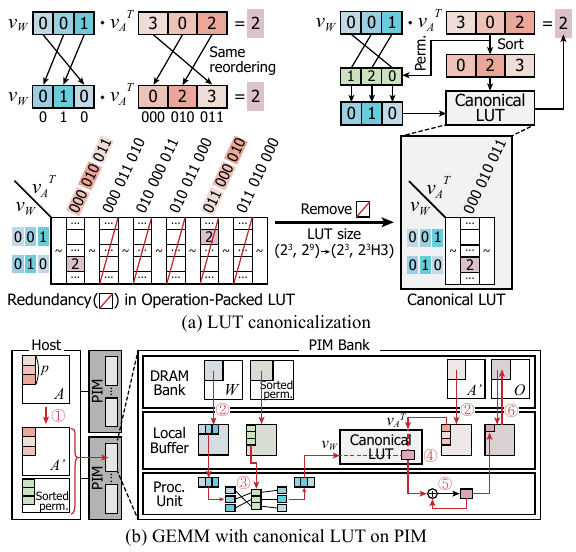}
    \caption{
    \Sort with the execution flow within a PIM architecture. (a) \sort eliminates the redundant entries included in an operation-packed LUT. (b) To access the \lut, the weight needs to be reordered. 
    }
    \label{fig:scheme1}
\end{figure}

\subsection{\Sort}
\label{sec:design:sort}
As discussed in \cref{sec:prelim}, the size of operation-packed LUT grows exponentially with respect to the bitwidths and the packing degree $p$.
{
Thus, if the LUT size can be reduced, this not only contributes to capacity savings but also results in performance gains by potentially using larger $p$ within the capacity budget.
}
Fortunately, operation-packed LUTs exhibit a high degree of redundancy in their entries. 
We exploit this by proposing \emph{\sort}, a method that significantly reduces the LUT size.

{
The main idea behind the \sort is that the operation-packed LUT has much internal redundancy. 
For example, it stores $a\cdot x + b\cdot y$ and $ b\cdot y + a\cdot x$ separately, which causes redundancy.
\cref{fig:scheme1}(a) illustrates the \sort method, with an example of $p=3, b_w=1$, and $b_a=3$.
In the example, the three weight values are multiplied by the three activation values. 
For this, the LUT is indexed with weight vector [0~0~1] and activation vector [011 000 010] to obtain the result.
However, a key observation is that the result remains invariant under any joint permutation of weights and activations.
For example, as shown in \cref{fig:scheme1}(a) the result value 2 can be obtained with weight vector [0~1~0] and activation vector [000 010 011], since both the weight and activation vectors are permuted in the same order from their original forms.
This results in a large number of functionally duplicate LUT entries.

To eliminate this redundancy, \sort enforces a fixed ordering on the activation vector\footnote{In principle, the same strategy could be applied to weights. However, fixing the activation order is generally more beneficial, as quantized DNNs typically use higher bitwidths for activations than for weights—that is, $b_a \geq b_w$.}. 
We simply use sorting to canonicalize the activation vector.
Due to this structure, 
accessing \sort\ requires sorting the input activations to determine the LUT access index. The input weights must then be reordered accordingly to match the sorted activations, a process we refer to as weight \textbf{reordering} with respect to activations.
}
This canonicalization ensures that only one representative input permutation is stored for each unique MAC combination, thereby reducing the size of the LUT substantially.

For comparison, the number of columns in the original LUT is $2^{p\cdot b_a}$.
On the other hand, the number of columns of the \lut is equal to the number of multisets:
\begin{align}
    \textit{\small canonical\_lut\_columns} = {}_{2^{b_{a}}}H_{p} = {\scriptstyle \binom{2^{b_a}+p-1}{p}},
\end{align}
for the activation bitwidth $b_a$ and packing degree $p$.
Because the growth rate is a high-order polynomial in $p$ at small fixed $b_a$, this grows slower than the original LUTs and therefore allows for selecting larger $p$ values. 
For example, when $b_a=1$, the reduction rate of the LUT size is 12.4$\times$ at $p=4$, which increases to 611.1$\times$ at $p=7$ (the largest $p$ we use in experiments), achieving drastic LUT size reduction.

A detailed procedure of \sort during DNN inference is illustrated in~\cref{fig:scheme1}(b).
\circled{1} The activation values are first quantized on the host, then $p$ activations are sorted and packed into a single activation vector (referred to as a canonicalized vector).
This canonicalized activation vector, along with its sorted permutation, is transferred to the PIM device.
\circled{2} To perform matrix multiplication, the processing unit inside the PIM loads the packed weights, the sorted permutation, and the canonicalized activation.
\circled{3} The packed weights are then unpacked, reordered according to the activation’s sorted permutation, and repacked to form the canonicalized weight vector, which serves as the row index of the LUT.
\circled{4} The canonicalized activation vector is used as the column index to access the LUT.
\circled{5} The resulting partial inner product value is accumulated within the processing unit, and \circled{6} once all computations for matrix multiplication are aggregated, the final output is written back to the memory bank.
As \sort can reduce the operation-packed LUT size, thereby enabling a larger $p$, the number of operations packed together can be increased (i.e., larger $p$).
This results in fewer LUT lookups, enhancing the overall system performance.


\begin{figure}
    \centering
    \includegraphics[width=0.92\columnwidth]{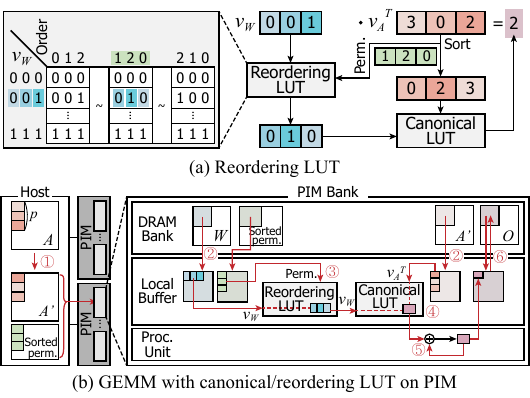}
    \caption{
     Illustration of \Reorder design. (a) Index reordering is replaced with a single \reorder lookup. (b) Execution flow with \reorder and \lut while processing GEMM.
    }
    \label{fig:scheme2}
\end{figure}

\subsection{\Reorder}
\label{sec:design:reorder}
Although \sort greatly reduces the LUT size and allows the use of a larger $p$, it comes at a large cost of weight index reordering.
As depicted in \cref{fig:scheme1}(a), each weight index has to be reordered according to the permutation order of the sorted activations. 
{In the figure, the original activations are [3 0 2] and the weights are [0 0 1]. 
Because the canonical form of the activations are [0 2 3], we permute the weights in the same manner into [0 1 0].} 
This bitwise operation adds a non-negligible overhead for PIM, either by incurring extra computation cycles for the general-purpose core or mandating an additional permutation logic in the processing unit.

To alleviate the overhead of runtime reordering, we introduce an auxiliary lightweight LUT called \reorder, which handles the reordering of weight vectors, as illustrated in~\cref{fig:scheme2}(a).
Instead of reordering the weight vector on the fly using the in-memory processing unit, \reorder stores the precomputed result in the LUT for each input case.
This LUT is indexed by the sorted permutation of the activation vector (as the column index) and the packed weight vector (as the row index).
Each entry in \reorder returns the reordered weight vector aligned to \sort.
As a result, \reorder offloads the reordering process---including unpacking, permutation, and repacking---into a single LUT lookup, significantly reducing runtime computation overhead.

\cref{fig:scheme2}(b) illustrates the execution procedure with the integration of \reorder.
Compared to \cref{fig:scheme1}(b), the key difference lies in step~\circled{3}, where the weight reordering is no longer performed by the in-memory processing unit.
Instead, it is replaced by a single lookup to the \reorder.
The rest of the execution flow remains unchanged.
Given the limited compute capability of processing units embedded in memory chips, this transition effectively alleviates a performance bottleneck, at the cost of slightly increased LUT capacity.

\begin{figure}
    \centering
    \includegraphics[width=\columnwidth]{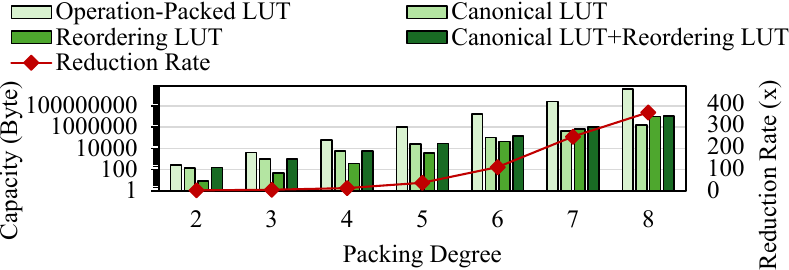}
    \caption{
    Capacity requirement for LUTs, varying packing degrees $p$ for W1A3 configuration. A red line indicates the total reduction rate of \thiswork compared to the operation-packed LUT. 
    }
    \label{fig:lut_cap}
\end{figure}

\cref{fig:lut_cap} shows the capacity requirements of the operation-packed LUT, \lut, and \reorder across various packing degrees at $b_{w}=1$ and $b_{a}=3$.
As \reorder contains all possible permutations of the $p$ activations, the number of columns in this LUT is $p!$.
Even though this adds some capacity overhead, the reduction in capacity achieved by \sort surpasses the overhead introduced by \reorder, achieving the total reduction rate by $1.68\times$ to $358\times$.
As a result, higher packing degrees become feasible compared to traditional operation-packed LUT designs, significantly improving LUT lookup efficiency.

\begin{figure}
    \centering
    \includegraphics[width=\columnwidth]{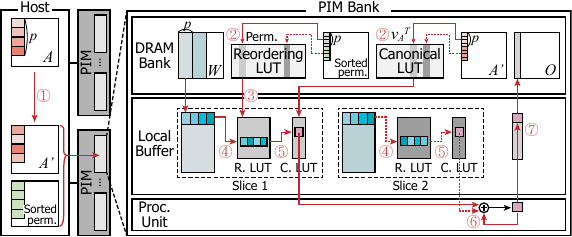}
    \caption{
    Procedure of \Dataflow. The LUTs are sliced into columns, and streamed to the local buffers for reuse.
    }
    \label{fig:scheme3}
\end{figure}

\subsection{\Dataflow}
\label{sec:design:dataflow}
While \sort and \reorder significantly reduce LUT size by eliminating redundant entries, their effectiveness is still limited by the local buffer size. 
This is because the maximum value of the packing degree $p$ remains constrained by the largest \lut.

{
To alleviate this limitation and better exploit the memory hierarchy of DRAM banks and the local buffer, 
we propose \dataflow, a technique that allows LUTs larger than the local buffer in DRAM banks while loading 
a subset of the LUT into the faster accessible local buffer.
For multiplying the target matrices, we first choose to keep the $k$ set of $p$ activations and multiply them by a series of weight values (the colored vertical strips of the $W$ matrix within the DRAM bank in \cref{fig:scheme3}), analogous to an input-stationary dataflow.
Because we fix the activations during the process, the accessed portion of both the canonical LUT and reordering LUT is confined to $k$ columns, which we refer to as \emph{slice}s.
Once the $k$ slices are loaded in the local buffer, the weight parameters are multiplied by accessing the corresponding rows from the slices.
This ensures that a slice is reused many times before advancing to the next activation values.


}

With \dataflow, the overall execution flow applying all the components (i.e., \sort and \reorder) is depicted in \cref{fig:scheme3}.
\circled{1} As in \cref{fig:scheme1} and \cref{fig:scheme2}, the host makes canonicalized activation vectors with sorted permutation, and sends them to the PIM devices.
\circled{2} Then, we first select $p\cdot k$ activations, where an integer $k$ denotes the number of columns of the \lut to be accessed.
For instance, in the case of $k=2$ shown in \cref{fig:scheme3}, two canonicalized activation vectors are chosen.
For each selected activation vector, a column slices from \lut are chosen by the canonicalized activation vector, while another pair of column slices from \reorder is chosen by the corresponding sorted permutation to be loaded into the local buffer.
\circled{3} The corresponding LUT slices are streamed from DRAM into the local buffer. 
This process involves loading $k$ columns from both the \lut and the \reorder.
\circled{4} The packed weight vectors are then used to index entries in the buffered \reorder column slices, producing canonical forms of the weight vectors.
\circled{5} These canonical weight vectors are then used to index the buffered \lut column slices.
\circled{6} Once these columns are loaded, we perform MAC operations across multiple weight vectors, reusing the buffered LUT column entries.
Each row access yields a partial output, and the reuse of LUT slices across weight rows enables efficient computation with minimal memory traffic.
\circled{7} Finally, when the output values are computed, the results are written back to the DRAM bank.

This reuse pattern resembles input-stationary dataflows in DNN accelerators, but with a key distinction: instead of holding activations stationary, \dataflow holds a slice of the LUT stationary while the weight matrix is read.
This approach is tailored specifically to LUT-based inference and is designed to maximize the reuse of the \lut under buffer size constraints.

By doing so, \dataflow significantly extends the effective packing degree beyond what buffer-sized LUTs can support, while avoiding frequent full-LUT reloads. 
It strikes a balance between leveraging DRAM’s large capacity and the buffer’s high bandwidth, enabling high-throughput, low-overhead inference with operation-packed LUTs.






\subsection{Performance Model}
\label{sec:design:model}
In this subsection, we present a first-order performance model to (1) determine the optimal packing degree $p^*$, and (2) identify when \dataflow outperforms a buffer-resident \lut implemented with \sort and \reorder.
We focus solely on the time consumed by LUT accesses. While weights, activations, and outputs also affect execution time, their contribution is marginal with respect to changes in $p$ or the choice of lookup strategy.

We model the execution time of matrix multiplication between a weight matrix tile $W\in\mathbb{Z}^{M\times K}$ and an activation matrix tile $A\in\mathbb{Z}^{K\times N}$.
The performance is determined by
\begin{enumerate}
    \item Streaming the relevant LUT slice from DRAM bank to the local buffer.
    \item Looking up the \reorder for creating the canonical weight vector.
    \item Looking up the \lut on the local buffer for the MAC operations.
\end{enumerate}
 Because a LUT slice contains $2^{b_wp}$ entries, the total execution time $T$ is
\begin{align}
T=2^{b_wp}\cdot\frac{KN}{p}\cdot L_D + \frac{MKN}{p}\cdot L_{local}, \label{eq:perf_slice}
\end{align}
where $L_D$ and $L_{local}$ are profiled constants. $L_D$ denotes the latency to load a single \lut entry and a \reorder entry from DRAM bank to the local buffer.
\\
$L_{local}\triangleq L_{\reorder}+L_{\lut}$ comprises one access to the \lut and one to the \reorder, along with additional accumulation for the partial output.
From this, the optimal packing degree $p^*$ is given by 
\begin{align}
            p^{*} = \mathop{\arg\min}\limits_{1 \leq p \leq p_{{}_{DRAM}}} \left\{ \frac{1}{p}\left( 2^{b_{w}p}\cdot{L_D} + M\cdot L_{local} \right) \right\}. \label{eq:pstar} 
\end{align}
With small $b_w$ (slower LUT growth) or large $M$ (more slice reuse), a larger $p^*$ is favored, potentially up to $p_{{}_{DRAM}}$ that fits the entire DRAM bank.

On the other hand, with a small $M$, not only is a smaller $p$ preferred, but a fully buffer-resident LUT may yield lower total cost.
In this case, the first term of \cref{eq:perf_slice} is removed. Using $p_{local}$ such that the LUT fits within the local buffer, the cost becomes:
\begin{align}
    T_{local}=\frac{MKN}{p_{local}}\cdot L_{local}. \label{eq:perf_local}
\end{align}
Thus, $T_{local}$ is shorter than $T$ when:
\begin{align}
\frac{1}{p_{local}}\cdot M\cdot L_{local}      < \frac{1}{p^*}(2^{b_{w}p^*}\cdot{L_D} + M\cdot L_{local}).      
\end{align}
Solving for $M$, we obtain
\begin{align}
    M < 2^{b_wp^*}\cdot\frac{ L_D}{L_{local}}\cdot\frac{p_{local}}{p^*-p_{local}}.
    \label{eq:decision}
\end{align}
{This indicates that \dataflow becomes beneficial as $M$ increases.}
Meanwhile, the condition aligns with the insight, such that the break-even point for $M$ increases with 1) larger $b_w$ (LUT size grows faster), 2) larger bandwidth gap between the DRAM and the local buffer (accessing LUT slice from DRAM is more costly), and 3) smaller gap between $p^*$ and $p_{local}$ (the gain on computational throughput is small).
Since the feasible range of $p^*$ is small (typically $p_{{}_{DRAM}}<10$ for 64 MB DRAM bank), we simply test all $p\leq p_{{}_{DRAM}}$ values on \cref{eq:perf_slice} and \cref{eq:decision} to automatically select the packing degree $p^*$ and determine whether to apply \dataflow or instead store the LUT entirely in the buffer.

\begin{figure}
    \centering
    \includegraphics[width=.95\columnwidth]{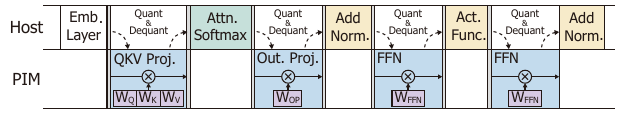}
    \caption{
    Transformer layer execution flow of \thiswork.
    }
    \label{fig:dnn_flow}
\end{figure}

\section{Implementation}
\label{sec:impl}
\subsection{Real-System Implementation on UPMEM}
\label{sec:impl:upmem}
We implement and evaluate our system on a server equipped with 32 ranks of UPMEM~\cite{truepim} DIMMs.
An UPMEM DIMM has 64 DRAM banks per rank\footnote{Eight banks per rank, where each bank is further split into eight pieces to be interleaved into 8 chips. We refer to each piece as a bank for simplicity in the context of UPMEM.}, where each bank contains a 64 MB DRAM array, a 64 KB SRAM-based local buffer, and an in-order general-purpose processor.
On the system, \thiswork was implemented using UPMEM SDK 2023.2.0~\cite{upmem_sdk}.

On the DRAM bank and the SRAM local buffer, we devote approximately half the capacity to the LUTs.
The remaining capacity is spent to store weights, input activations, outputs, and other miscellaneous data. 
Thus, within the 64 MB DRAM bank, $p_{{}_{DRAM}}$ is around 8 for $b_a=1$ and $b_w=3$.
Within the 64 KB local buffer, $p_{local}$ is around 5 for the same bitwidth. 
Without the \sort technique, $p_{{}_{DRAM}}$ and $p_{local}$ decreases to 6 and 3, respectively.

At initialization, the performance model from \cref{sec:design:model} is computed on the host side using the matrix dimensions given to determine $p^*$ and whether to use \dataflow.
Afterwards, the LUT is constructed according to the parameters and is broadcast to all banks.

Once the input activations are prepared, they are tiled to fit within the banks.
Each activation tile is then divided into vectors containing $p^*$ elements and sorted into the canonical order.
These vectors, along with their sorted permutations, are distributed to the banks. 
Once the multiplication is complete at each bank, they are gathered at the host and combined to form the output matrix.

\subsection{DNN Workflow}
\label{sec:impl:dnn}
We mainly implement and evaluate \thiswork on transformer-based DNN models in various fields.
As illustrated in \cref{fig:dnn_flow}, the PIM-enabled banks perform all matrix multiplication operations (i.e., QKV projection, output projection, and FFN), while the host processor handles the rest, especially the fp32 operations such as softmax, normalization, and GELU.

Among the 2048 banks, data parallelism and context parallelism are used to split the workload. 
Before and after the PIM operations, the host processor performs quantization and dequantization to allow low-bit integer arithmetic.
Note that this does not incur additional PIM-host communication, because performing consecutive matrix multiplications involves inter-bank communication, which has to travel through the host processor even without the quantization~\cite{pidcomm}.



\section{Evaluation}
\subsection{Experimental Setup}

\noindent\textbf{Environment.} We implemented and evaluated all baselines and \thiswork{} on a real PIM-enabled UPMEM system \cite{truepim}.
The system includes an Intel Xeon Gold 5215 CPU and 32 ranks of PIM-enabled DIMMs.
In UPMEM architecture, each bank is equipped with its own processing element (PE), allowing for the utilization of up to 2048 PEs.
To further explore the potential of extending the accelerator-driven design of other PIM architectures~\cite{hbmpim}, we also employed an in-house cycle-accurate simulator to evaluate the performance benefits that could be achieved on a custom hardware (\cref{sec:discussion:hbmpim}).

\noindent\textbf{Baselines.}  
We extensively compare the performance of \thiswork against two LUT-based solutions and two product-quantization (PQ)-based solutions.  
For the LUT-based solutions, we evaluate performance and robustness across various conditions and GEMM operations, while for the PQ-based solutions, we additionally analyze model accuracy.  

\vspace{0.5em}
\noindent{\textbf{Baselines for direct comparison:}}
\begin{itemize}[leftmargin=*]
\item{\textbf{Naive PIM}}: Without the use of LUTs, a conventional PIM would utilize its attached arithmetic units to perform matrix multiplications. With Naive PIM, we implement the matrix multiplication kernels using the in-order processors, utilizing the attached int8 multipliers. 

\item \textbf{LUT Tensor Core (LTC)}~\cite{luttensorcore}: LTC is a software-hardware co-design optimized for low-bit LLM inference using LUT-based mixed-precision GEMM (mpGEMM). This introduced LUT compression techniques and tiling tailored to mixed-precision matrix multiplication. Although the original LTC design is implemented and evaluated using Verilog, we faithfully adapted its core ideas to our environment.
Additionally, since T-MAC~\cite{tmac} shares similar core principles when adopting PIM, we focus on LTC.
\end{itemize}

\vspace{0.5em}

\noindent{\textbf{\thiswork design points:}}
\begin{itemize}[leftmargin=*]
    \item{\textbf{Operation-Packed LUT (OP)}}: 
Operation-packed LUT serves as the basic starting point for the proposed method, where each LUT entry stores precomputed dot products for multiple activation patterns, enabling higher arithmetic intensity per memory access. 
We use the simple operation-packed LUT sized for the local buffer as a baseline to isolate and evaluate the incremental performance benefits introduced by \thiswork.
 \item{\textbf{\Sort (OP+LC)}} applies \sort on top of OP to enable larger packing degrees that fit in the local buffer.  
 \item{\textbf{\Reorder (OP+LC+RC)}} further applies \reorder to alleviate the computation overhead for weight reordering.
 \item{\textbf{\Dataflow (OP+LC+RC+SS (\thiswork))}} represents the final proposed design, with the addition of \dataflow for utilizing the DRAM banks.
\end{itemize}

\vspace{0.5em}
\noindent{\textbf{PQ-based LUT design baselines:}}
\begin{itemize}[leftmargin=*]
    \item \textbf{PIM-DL}~\cite{pimdl}: PIM-DL is the state-of-the-art quantized DNN framework for PIM architectures. It partitions the LUT, which stores partial MAC results of the weight and activation matrices, across the PEs, enabling independent LUT additions to maximize parallelism.  
    \item \textbf{LUT-DLA}~\cite{lutdla}: LUT-DLA develops efficient accelerators for PQ and introduces a PQ training method to preserve accuracy. To reduce the overhead of centroid–activation similarity computations, it supports both L1 and L2 (Euclidean) distance metrics. We refer to the two variants as LUT-DLA (L1) and LUT-DLA (L2), respectively. 
\end{itemize}

\noindent\textbf{Workloads.} We benchmarked various transformer-based deep learning models~\cite{transformer} by quantizing them with diverse quantization setups. 
We tested three representative transformer-based deep learning models.
\begin{itemize}[leftmargin=*]
    \item \textbf{Encoder-only language model (BERT)~\cite{bert}}: We adopted this model because it proposes the base structure of the encoder-only language models. We select BERT-base model.
    \item \textbf{Decoder-only language model (OPT)~\cite{opt}}: We selected this model since recent decoder-only models employ a similar structure to it. 
    We choose OPT with a model size of 125M parameters.
    \item \textbf{Vision Transformer (ViT)~\cite{vit}}: This popular vision model regards patches of images as tokens. We selected the ViT-Base model with 86M parameters.
\end{itemize}

\noindent\textbf{Quantization.}
We evaluated four activation/weight bit-width settings, denoted as \textbf{W$x$A$y$}: covering a range of quantization options for both activations and weights. 
Since the quantization method (e.g., how to round to an integer) hugely affects model accuracy, we adopt the existing prior art for obtaining the low-bit quantized weights: W1A3~\cite{binarybert}, W1A4~\cite{binarybert}, W2A2~\cite{kdlsq}, and W4A4~\cite{kdlsq} are used for BERT, W2A2~\cite{qvit} and W4A4~\cite{qvit} for ViT, and W4A4~\cite{omniquant} for OPT.

\noindent\textbf{Datasets.}
For language models (BERT and OPT) we utilize the GLUE benchmark, which includes four representative tasks: QNLI, QQP, STS-B, and SST-2. 
QQP and QNLI determine the semantic relationship between sentence pairs, 
STS-B measures sentence-level semantic similarity, and SST-2 classifies movie reviews as positive or negative.  
We also set a maximum input length to 128 for these tasks.
For ViT, we employ the Imagenet \cite{imagenet} dataset, a popular benchmark for assessing vision tasks.

\begin{figure}
    \centering
    \includegraphics[width=\columnwidth]{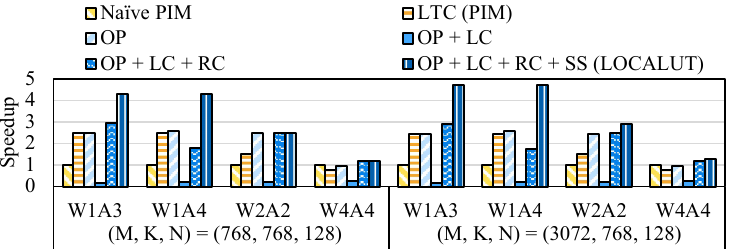}
    \caption{
    Performance comparison for GEMM operation.
    }
    \label{fig:eval_speedup_gemm}
\end{figure}

\subsection{Performance of GEMM Operations} \label{eval:gemm}
In \cref{fig:eval_speedup_gemm}, we evaluate the performance of the GEMM operation across baselines and \thiswork.
The configurations are denoted as (M, K, N), which corresponds to multiplying an M$\times$K matrix with a K$\times$N matrix to produce an M$\times$N output.
To reflect representative DNN workloads, we use two configurations---(768, 768, 128) and (3072, 768, 128)---commonly used in the BERT, ViT, and OPT models.

Overall, \thiswork exhibits a geometric mean speedup of 2.87$\times$ over Naive PIM, and 1.77$\times$ speedup compared to LTC, respectively.
Naive PIM suffers from the limited computational power and fails to effectively leverage the high internal bandwidth. 
LTC also faces performance bottlenecks, as its LUTs often contain redundant entries, leading to low LUT packing degrees. This sometimes leads to inferior performance compared to the naive PIM method.
%
In contrast, operation-packed LUT (OP) already performs slightly better, by allowing multiple operations in a single lookup. 
From the optimizations, \thiswork gains significant speedup, achieving speedup of up to 4.73$\times$ over Naive PIM and 1.93$\times$ over LTC. 

We observe that all three additional optimizations are essential for the speedup. 
Although \sort (OP+LC) allows using larger LUTs with higher $p$, its performance drops significantly from the added ordering overhead at the processing unit.
Further applying \reorder (OP+LC+RC) alleviates this overhead with the introduction of an auxiliary LUT. 
Finally, \dataflow exploits the large capacity of the DRAM banks for $p$, leading to the final speedup of \thiswork.
For W2A2 and W4A4, a high volume of LUT loading reduces performance, especially in the matrix (768, 768, 128) where RC is not utilized according to the performance model.
However, RC benefits from larger $p$ by reusing LUT columns as M increases to 3072.
Additionally, we observe that the speedup of \thiswork is more substantial in lower-bit quantization configurations.
This is because lower-bit precisions result in smaller LUTs, leaving more room to pack the LUTs to a higher degree. This aligns well with the trend of movements toward extremely low-bit quantization.

\begin{figure}[t]
    \centering
    \includegraphics[width=\columnwidth]{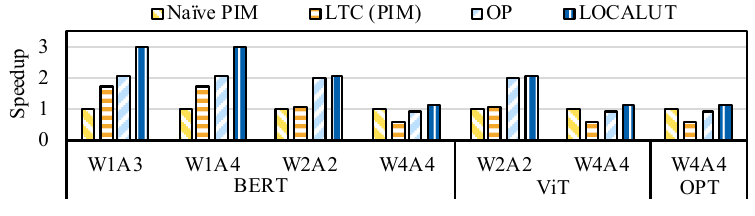}
    \caption{
    Performance comparison among different DNN models.
    }
    \label{fig:eval_speedup_various_models}
\end{figure}

\subsection{Performance of Representative DNN Workloads}
In \cref{fig:eval_speedup_various_models}, we further measure the end-to-end speedup of \thiswork across different DNN models and quantization configurations.
The results indicate that \thiswork consistently outperforms the baselines, achieving speedups of 1.77$\times$ over Naive PIM, and 1.82$\times$ over LTC. 
Comparison between OP and OP+LC+RC+SS further shows that the optimizations proposed in \cref{sec:design} add 22\% additional speedup.
As these models are composed mainly of GEMM operations, their performance trends closely align with those observed in the standalone GEMM evaluations presented in \cref{eval:gemm}.

\begin{figure}
    \centering
    \includegraphics[width=.87\columnwidth]{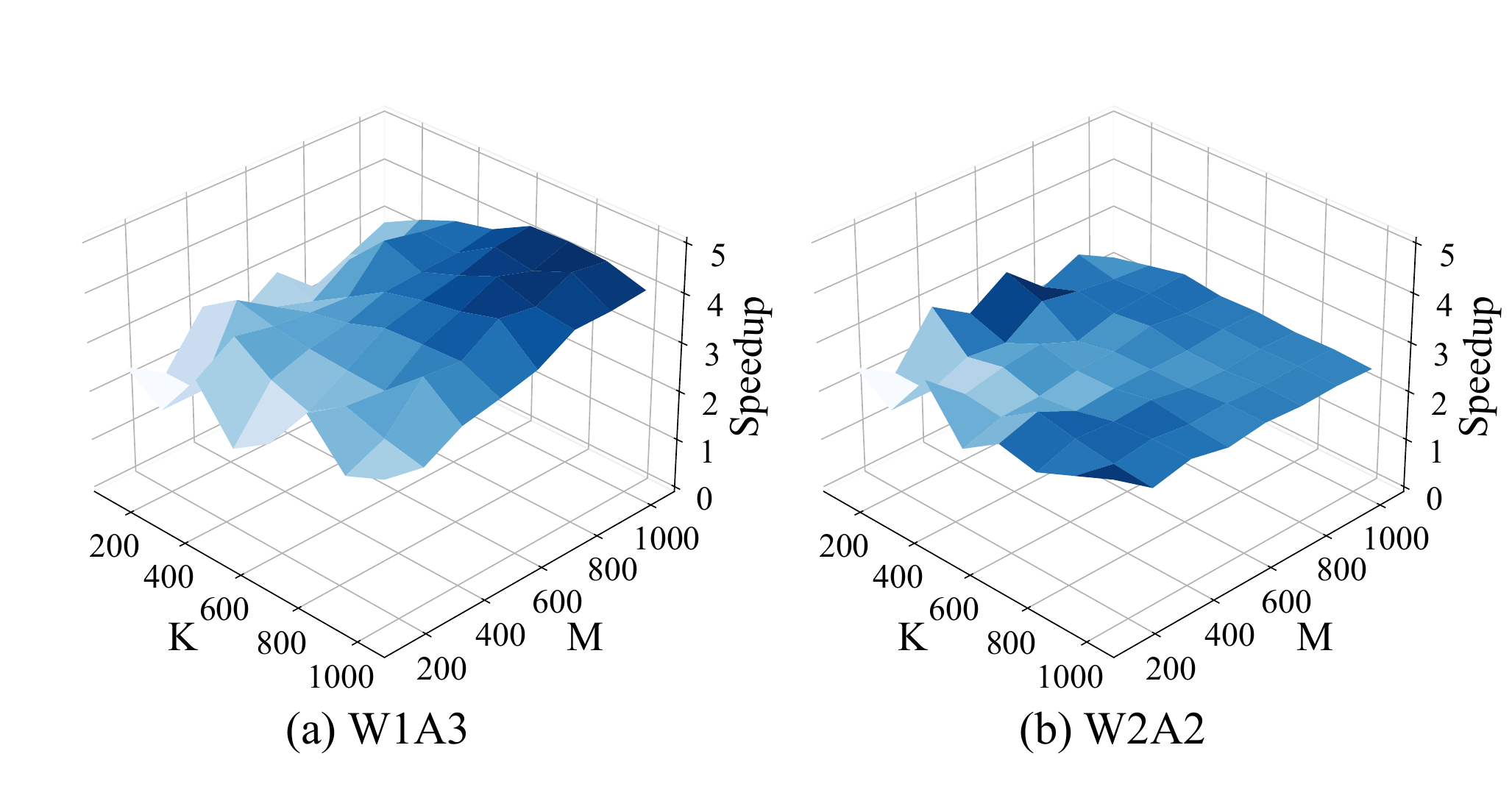}
    \caption{
       Sensitivity study of \thiswork\ with varying weight matrix dimensions. (N = 128)
    }
    \label{fig:eval_matix_size_sensitivity}
\end{figure}

\subsection{Sensitivity Study}
\label{sec:sensi}

\textbf{Matrix size}. 
\cref{fig:eval_matix_size_sensitivity} presents the matrix size sensitivity of \thiswork, normalized to the Naive PIM baseline. 
We evaluate two representative quantization settings, \textbf{W1A3} and \textbf{W2A2}, to cover both extreme low-bit and moderate precisions.
The weight matrix size is varied from (128, 128) to (1024, 1024), while the activation width is fixed at N $=$ 32.

As shown in both \cref{fig:eval_matix_size_sensitivity}(a) and (b), \thiswork consistently outperforms the baseline across all tested matrix dimensions.
We observe a geometric mean speedup of $2.86\times$ under both quantization settings, demonstrating the robustness of our approach regardless of matrix size.  
These results validate that the LUT-centric optimizations in \thiswork generalize well across varying compute intensities and memory footprints.


\begin{figure}
    \centering
    \includegraphics[width=\columnwidth]{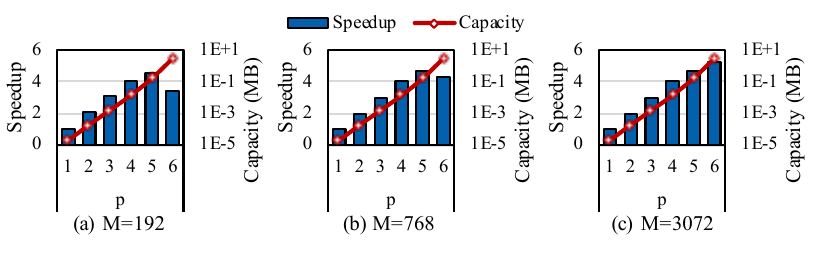}
    \caption{
    {Packing degree ($p$) sensitivity with varying M dimension (K$=$768, N$=$128, W2A2 precision). (a) M$=$192, (b) M$=$768, (c) M$=$3072.}
    }
    \label{fig:p_sense}
\end{figure}

{
\textbf{$\boldsymbol{p}$ sensitivity}.
\cref{fig:p_sense} demonstrates the tradeoff between performance and capacity when K$=$768, and N$=$128 for W2A2.
As the packing degree $p$ rises, the performance improves while the capacity increases.
An interesting observation is that performance improves as M increases at $p=6$.
This occurs because increasing the $p$ expands the size of the LUT slice, which introduces additional overhead.
Although capacity also grows with higher $p$, \thiswork leverages DRAM to store the entire LUT and uses slices of the LUT.
}

\begin{figure}
    \centering
    \includegraphics[width=\columnwidth]{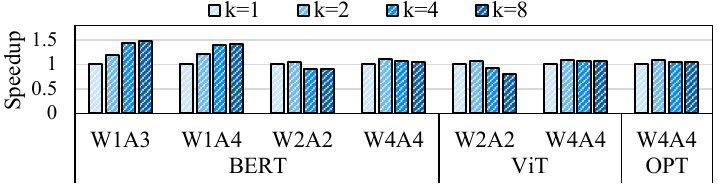}
    \caption{
    {Sensitivity to the $k$ values.}
    }
    \label{fig:rev_eval_k_sensitivity}
\end{figure}

{
\textbf{$\boldsymbol{k}$ sensitivity}.
We vary the number of slices of the \lut $k$ from 1 to 8 across different models and bit-widths, as shown in \cref{fig:rev_eval_k_sensitivity}.
For each chosen $k$, we select the highest $p$ possible in the remaining memory space.
In principle, increasing $k$ results in higher performance, because keeping more slices in the local memory improves the data reuse, as exemplified by W1Ax configurations.
However, larger $k$ requires more space in the local memory, which could lead to a lower $p$ that fits in the remaining memory.
This is why increasing $k$ from 2 to 4 in W2A2 and W4A4 settings degrades the performance.
Because of insufficient space left in the local memory, lower $p$ value is used, which leads to a slowdown.

}

\begin{figure}
    \centering
    \includegraphics[width=0.95\columnwidth]{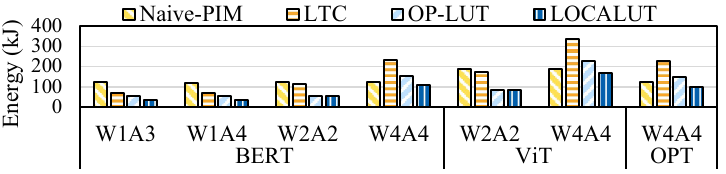}
    \caption{
    {
    Energy comparison between baseline methods and \thiswork. Three models (BERT, ViT, OPT) with different bitwidth configurations.
    }
    }
    \label{fig:rev_eval_pim_energy}
\end{figure}

{
\subsection{Energy Consumption}
\cref{fig:rev_eval_pim_energy} compares the energy consumption of \thiswork against three baseline methods across BERT, ViT, and OPT models with various bitwidth configurations. \thiswork consistently demonstrates superior energy efficiency compared to all baselines. For W1Ax configurations, \thiswork shows 3.37$\times$ energy reduction over Naive-PIM and 1.88$\times$ energy reduction over LTC, demonstrating the effectiveness of exploiting high packing degrees and reducing computation through efficient LUT memory management via \sort.
For W2A2, the energy efficiency of \thiswork is better than Naive-PIM and LTC, but is on par with OP-LUT.
This is because \thiswork has extra sorting overheads that offset the reduced LUT lookups.
%
Finally, with W4A4, LTC and OP-LUT underperform Naive-PIM due to the low packing degree.
In contrast, \thiswork achieves 1.16$\times$ energy efficiency than Naive-PIM, demonstrating its effectiveness.
}

\subsection{Performance Comparison of LUT-based Methods with Product Quantization}
\begin{figure}
    \centering
    \includegraphics[width=.85\columnwidth]{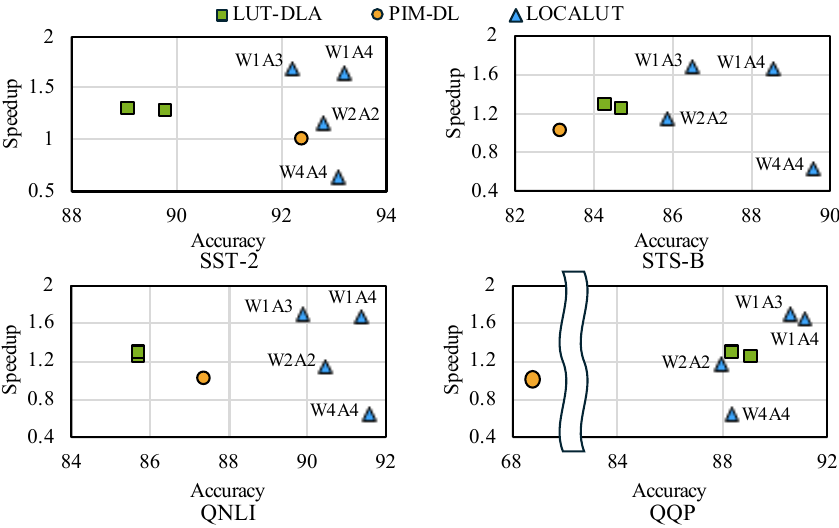}
    \caption{
       Performance comparison with existing product quantization methods. 
    }
\label{fig:eval_pq_scatter_plot}
\end{figure}

There have been several proposals to use LUTs in conjunction with product quantization~\cite{pimdl, lutdla}.
Unfortunately, comparing them directly with \thiswork in execution time would be unfair, because they do not adhere to fixed bitwidths. 
%
Instead, in \cref{fig:eval_pq_scatter_plot}, we jointly plot the execution time and accuracy on the BERT-base model to enable a fair comparison.
For PIM-DL~\cite{pimdl}, we reproduced the accuracy results using the official implementation.
For LUT-DLA~\cite{lutdla}, we use the reported accuracy due to the lack of public implementation. 
As the downstream tasks do not affect the model structure, their speedups remain identical over all benchmarks.
While the accuracy gap varies across benchmarks, it is evident that \thiswork provides a clear advantage in terms of speed and accuracy over 
the PQ-based LUT methods.

\subsection{Kernel Breakdown}
\label{sec:eval:breakdown}
\begin{figure}
    \centering
    \includegraphics[width=\columnwidth]{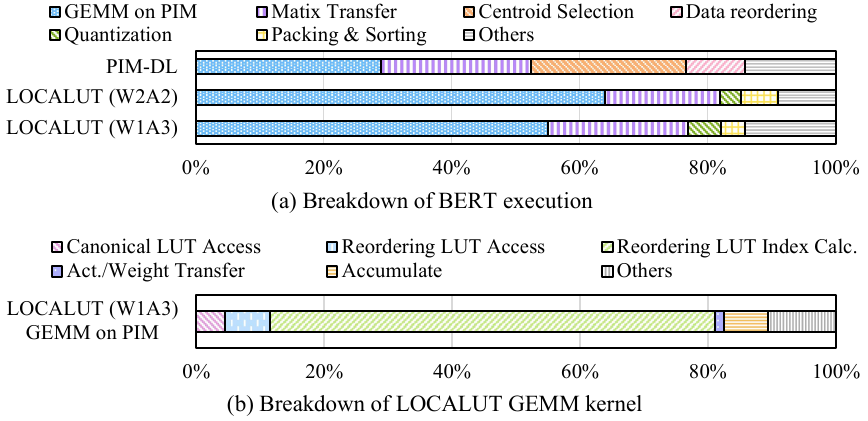} 
    \caption{
    {(a) Comparison of \thiswork and PIM-DL (b) Breakdown of \thiswork on the GEMM kernel.}}
    
    \label{fig:rev_eval_breakdown}
\end{figure}

\cref{fig:rev_eval_breakdown}(a) sheds some light on the reason for performance difference between \thiswork and the PQ-based approaches.
We compared PIM-DL with W2A2 and W1A3, which show comparable accuracy in SST-2 benchmark.
As depicted, PIM-DL consumes a smaller portion of time on the core GEMM operations on PIM. 
However, it exhibits a large overhead on the host processor instead, for finding the centroid for each value. While \thiswork also has overhead at the host processor for the quantization, this is more lightweight. 

{
\cref{fig:rev_eval_breakdown}(b) shows time breakdown for the GEMM kernel.
Counter-intuitively, LUT access only consumes a small portion, while reordering LUT index calculation dominates execution.
Additionally, the latency overhead caused by the reordering LUT access accounted for 6.9\% of the total kernel execution time.
While the reordering LUT itself reduces much of the index calculation, the processing unit's limited computational capability struggles with the remaining bitwise operations. 
We believe dedicated hardware accelerators or more powerful compute units could further address this bottleneck.
}

{
\subsection{Comparison with CPU and GPU}
}
\label{sec:eval:other_hw}

\begin{figure}
    \centering
    \includegraphics[width=\columnwidth]{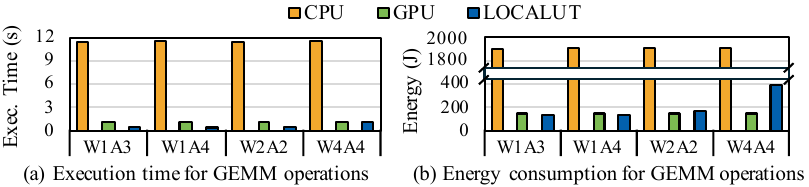}
    \caption{
    {Execution time and energy comparison with CPU and GPU across diverse bitwidths (M=12288, K=192, N=65536).}
    }
    \label{fig:rev_gpu}
\end{figure}

{
We further evaluate \thiswork against a conventional CPU- and GPU- based inference baselines. 
In \cref{fig:rev_gpu}, execution time and energy consumption are depicted for GEMM operation on multiple bitwidths.
For comparison, an Intel Xeon Gold 5215 CPU and an NVIDIA 2080Ti GPU are used.
While \thiswork consistently outperforms the CPU, the advantage over the GPU varies depending on the bitwidth.
Aligned with the results from \cref{sec:eval:breakdown}, the speedup is maintained for low bitwidths, where degradation occurs on the relatively higher W4A4 bitwidth.
This partially comes from the fact that UPMEM consumes much time and energy from the slow arithmetic units and frequent CPU-PIM communication, 
which prohibits us from drawing the full potential of PIM. 
Thus, we believe there is much room for improvement.

}

\subsection{Cost Model Validation}

To validate the performance model presented in~\cref{sec:design:model}, we compare the model-predicted execution times against actual system measurements across varying packing degrees.
To determine the values of parameters $L_D$ and $L_{\text{local}}$ in the performance model on our experimental setting, we characterized the behavior of the UPMEM platform~\cite{truepim} used in our evaluation. 
Specifically, data transfer from the DRAM bank to the local buffer occurs at a rate of \(0.5\,\mathrm{B/cycle}\), with each processor operating at \(350\,\mathrm{MHz}\). 
Considering a three-stage pipelined access, this results in \(L_D = 1.36 \times 10^{-9}\,\mathrm{s}\).
Additionally, lookup operations for \lut and \reorder with accumulation consist of 12 instructions, resulting in \(L_{\text{local}} = 3.27 \times 10^{-8}\,\mathrm{s}\).
We evaluate two precision configurations, W2A2 and W4A4, across two matrix sizes.

\begin{figure}
    \centering
    \includegraphics[width=\columnwidth]{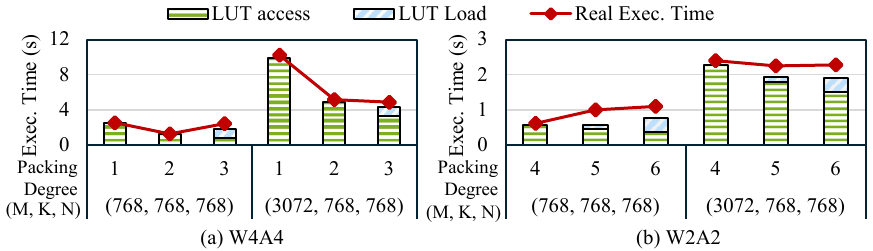}
    \caption{
    Validating the cost model's predicted execution time against real system results across different matrix sizes.
    }
\label{fig:eval_cost_model}
\end{figure}

As shown in~\cref{fig:eval_cost_model}(a), for W4A4, the model correctly identifies the packing size in both matrix sizes (two and three, respectively).
For W4A4, a maximum packing degree of two fits in the local buffer, which makes a packing degree of two outperform one in both matrix sizes.
However, increasing the packing degree to three introduces \dataflow, hence requiring LUT loading time.
Here, the reuse potential of the loaded LUT slice, which is determined by the weight matrix dimension, has a significant impact on performance.
It degrades for smaller matrices like \((768, 768, 768)\), but improves for larger ones such as \((3072, 768, 768)\), making higher packing degrees more beneficial for larger weight matrices.

For W2A2, while the model correctly determined a packing degree of five as optimal for the matrix \((3072, 768, 768)\), it slightly mispredicts on the matrix (\(768, 768, 768)\), selecting five instead of four. 
This is because the model does not consider all factors such as input value loading. 
However, the difference is small, and the model generally predicts correctly. 

{
\subsection{Performance in Various Scenarios}
}

\begin{figure}
    \centering
    \includegraphics[width=0.95\columnwidth]{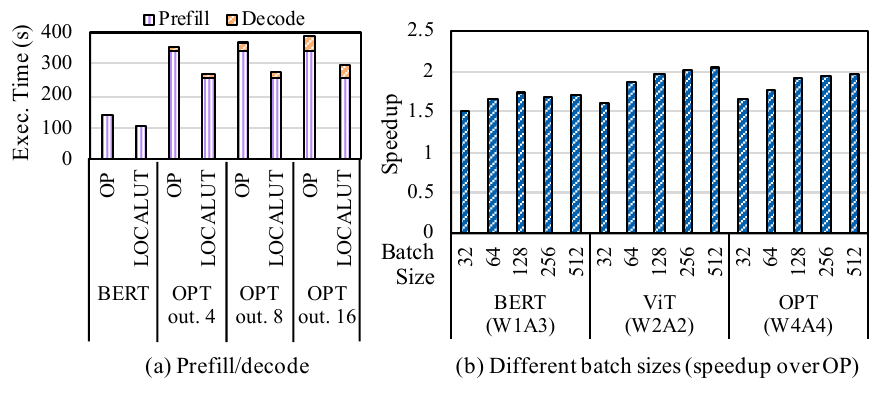}
    \caption{
    {
    Performance of \thiswork on real-world scenarios. 
    (a) Performance comparison of prefill-only model (BERT) and prefill$+$decode model (OPT).  
    (b) Evaluating \thiswork on varying batch sizes. 
    }
    }
    \label{fig:eval_prefill}
\end{figure}

{
In \cref{fig:eval_prefill}, We display \thiswork's performance in various scenarios.
Specifically, we measure the performance gain on prefill and decode phases of LLM inference, and also on different batch sizes to represent various input lengths.

We compare the execution time for prefill-only (BERT) and the prefill$+$decode (OPT) models, decomposing each into phases to show \thiswork's phase-wise performance gains.
We used W1A3 for BERT and W4A4 for OPT, evaluating OPT with various output lengths.
Results in \cref{fig:eval_prefill}(a) show that the prefill phase achieves a 1.34$\times$ speedup, while the decode phase achieves a 1.27$\times$ speedup, demonstrating \thiswork's performance benefits across phases.
%
\cref{fig:eval_prefill}(b) shows \thiswork's performance gains over OP across batch sizes 32-512.
\thiswork consistently achieves speedup, especially at high batch sizes by exploiting bank-level parallelism.
}

\begin{figure}
    \centering
    \includegraphics[width=0.99\columnwidth]{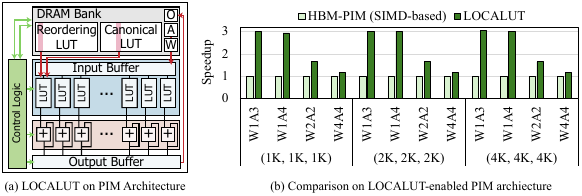}
    \caption{
    Comparison on \thiswork-enabled PIM architecture and HBM-PIM. (M, K, N) indicates the matrix size. 
    }
\label{fig:eval_hbm_pim}
\end{figure}

\vspace{2mm}
{
\subsection{\thiswork on Bank-level PIM.}
}
\label{sec:eval:hbmpim}

{
\textbf{Integration into Bank-level PIM.}
Since our LUT designs and the proposed \dataflow are not tied to a specific PIM architecture (i.e., UPMEM), they can be applied to re-design conventional parallel MAC-based bank-level PIM into a LUT-based design. 
As shown in \cref{fig:eval_hbm_pim}(a), rather than deploying SIMD processors near each bank, we leverage parallel \lut instances with \dataflow to perform MAC operations in memory.

Compared to the prior art, bank-level PIM design with SIMD processors~\cite{attacc}, our design replaces the 16-lane SIMD unit with sixteen 512B \lut units per bank.
We estimate the area overhead of the SRAM-based \lut using CACTI 7.0~\cite{cacti}.
We scale the results to a third-generation 10nm-class (1z-nm) DRAM process~\cite{hbm3_jssc}, considering that DRAM fabrication is approximately 10$\times$ less dense than logic processes at the same feature size~\cite{truepim}.
The arithmetic unit in~\cite{attacc} occupies $0.0592mm^2$ per bank, while the area overhead of our \lut-based design closely matches this, at $0.0591mm^2$ per bank.
{
Our design is also equipped with multiple accumulators and input/output buffers controlled by the control logic.
To enable LUT-based computation, we assume that the PIM instruction/command for MAC is modified to replace conventional MAC operations with LUT accesses.
Similar to other PIM architectures~\cite{aim, hbmpim, piccolo, bufcmp}, the control logic manages both input and output buffer accesses within each memory bank.

}

We implement a bank-level PIM design using SIMD processors~\cite{samsung-pim, attacc} and our LUT-based design on Ramulator 2.0~\cite{ramulator2}.
Considering the limited capacity of LUT per bank, we carefully tune $p$ and apply \sort and \dataflow.
As shown in \cref{fig:eval_hbm_pim}(b), \thiswork achieves 2.04$\times$ geometric mean speedup in matrix multiplication over the SIMD-based DRAM-PIM architecture across various sizes.
As the bitwidth increases, the required capacity for \lut grows, limiting the feasible packing degree.
Nevertheless, with W4A4 configuration, \thiswork still achieves a 1.17$\times$ speedup over the baseline PIM architecture.
}

\begin{figure}
    \centering
    \includegraphics[width=\columnwidth]{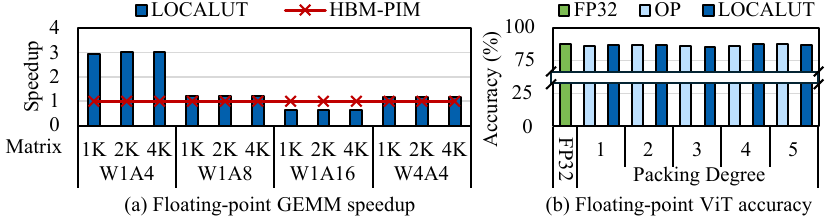}
    \caption{
    {Floating-point support in \thiswork. (a) GEMM speedup across different precisions (FP4, FP8, FP16) where matrix size indicates M=N=K (e.g., 1K = 1K×1K×1K). (b) ViT accuracy with varying packing degrees.}
    }
    \label{fig:eval_float}
\end{figure}

{
\textbf{Support for floating points.} 
\label{sec:eval:fp}
Using the architecture integrated into bank-level PIM, we extend \thiswork to support quantized floating-point operations in \cref{fig:eval_float}. 
Since the LUT entry count depends solely on input bitwidth rather than numerical format, \thiswork naturally generalizes to floating points. 
We evaluate \thiswork across varying precisions using three matrix configurations, benchmarked against HBM-PIM with the same floating point support in \cref{fig:eval_float}(a). 
For W1A4, W1A8, and W4A4 \thiswork attains up to 2.99$\times$, 1.22$\times$, and 1.17$\times$ speedups over HBM-PIM, respectively. 
However, W1A16 setting yields 0.62$\times$ geometric mean slowdown due to HBM-PIM's native FP16 hardware support.
Nevertheless, the overall results illustrate that \thiswork can deliver competitive performance on floating point operations. 

Moreover, we evaluated ViT's accuracy under the W4A4 setting using floating-point operations with (\thiswork) or without (OP) \reorder as shown in \cref{fig:eval_float}(b). 
A potential concern is that floating-point reordering may introduce numerical instability.
However, we find that \reorder produces negligible accuracy impact on ViT across packing degrees up to 5.
This robustness likely stems from structural similarity between \reorder and established tiling/reordering strategies in optimized GEMM/GEMV kernels, which preserve numerical stability. 
}

\vspace{2mm}
\section{Discussion}
\label{sec:discussion}

\vspace{2mm}

\subsection{Generality of \thiswork}
\label{sec:discussion:hbmpim}

{
In \cref{sec:eval:fp}, we explored applying \thiswork to existing hardware such as HBM-PIM~\cite{hbmpim}.
Replacing SIMD units with LUT-based logic slightly sacrifices generality for low-bit MAC operations.
However, many existing PIM products (e.g., HBM-PIM, AIM~\cite{aim}) also specialize in MAC operations, typically with single-precision support.
Hence, unlike the MAC-based bank-level PIMs, \thiswork offers some extra flexibility by supporting diverse precisions without hardware modification. 
Additionally, LUTs' reconfigurability allows supporting other operations (e.g., bitwise xor), provided they fit within the LUT capacity budget.
}

{
One promising direction is integrating \thiswork into GPU memory and compute units.
While simply attaching \thiswork-enabled DRAM to GPUs is possible, a tightly coupled architecture could embed LUT-based units within the GPU's arithmetic cores, leveraging GPU caches and scratchpad memory similar to local buffers storing LUT slices.
This opens new research opportunities, such as dynamic partitioning decisions between GPU-side and DRAM-side LUT processing.
}

\subsection{Limitations}
\label{sec:discussion:limitations}

{
\textbf{Capacity overhead.} 
\thiswork maximizes DRAM bank capacity utilization for performance gains.
However, this may limit applicability for models with large parameter sizes or in memory-constrained environments, especially when capacity is shared by multiple jobs.
Moreover, replicating LUTs across DRAM banks causes redundant memory usage.
Therefore, the LUT design and execution strategy may need to be tuned to balance performance and capacity efficiency within a constrained memory budget.
Although this capacity–performance tradeoff is a fundamental aspect of our approach, efficiently managing it remains an open challenge.
}



\textbf{Host processor dependency.}
Like many PIM systems, our current implementation offloads non-matrix operations (e.g., softmax, layer norm, GELU) to the host processor. 
However, we believe such overhead is inherent for most PIM designs. 
As supporting all general-purpose operations is costly for PIM, it's common to utilize host rather than solely relying on PIM~\cite{attacc, duplex, smartinfinity, gradpim, tensordimm}.
%
In addition, inter-unit communication is almost unavoidable for modern workloads, including DNNs and other large-scale parallel workloads \cite{megatron-lm, blink, sancus, optimus-cc, themis, pathweaver}, and for PIM, this is typically done through the host processor~\cite{pidcomm}, which is the hub for the memory channels.
Even though there are several attempts to enable direct communication between banks~\cite{rowclone, ndpbridge} or DIMMs~\cite{dimmlink}, it's still difficult to implement in practice. 
%

\textbf{{Low-bit quantization.}}
{
While \thiswork demonstrates significant improvements for DRAM-PIM-based inference, it is primarily constrained to low-bit quantized models.
As the bitwidth increases (e.g., $b_w > 4$), \lut and \reorder sizes grow substantially, degrading performance relative to arithmetic circuit-based designs.
However, our target aligns with emerging trends that increasingly favor low-bit quantization.
Examples include Microsoft's near 1-bit LLM~\cite{bitnet}, Apple's 2-bit on-device models~\cite{gunter2024apple}, and widespread adoption of 4-bit representations~\cite{omniquant, llmfp4, nvfp4, amxfp4, 4bitquant}.
Such low-bit approaches are being actively explored to reduce model size and enable deployment on resource-constrained environments.

}

\section{Related Work}

\textbf{Processing-in-memory.}
With the rise of data-intensive workloads, PIM has gained attention as a promising solution. 
Specifically, near-bank processors~\cite{gradpim, truepim, newton, aim, hbmpim} are being considered as one of the practical directions for near-future PIM.
Recently, memory vendors have produced real prototypes~\cite{truepim, aim, hbmpim}.
UPMEM~\cite{truepim} launched the first commercially available PIM product, featuring DRAM DIMMs with a general-purpose RISC core per bank, enabling various applications~\cite{upmem_benchmark, pidjoin, pidcomm, spidjoin, pimdal, PyGim, swiftrl, simplepim, virtualpim, transpimlib, upimulator, upmem_casestudy, fala}. 
Samsung introduced HBM-PIM~\cite{hbmpim} with compute units tailored for simple reductions and matrix-vector multiplications, where AiM~\cite{aim} from Hynix proposed an application-specific PIM solution for AI. 
While these products successfully use high internal bandwidth of DRAM, the tight area and power constraints of memory chips limited their compute capabilities.

\textbf{Lookup table for efficient processing.}
As mentioned in \cref{sec:bg:lut}, LUTs are considered an efficient method for matrix multiplications, sometimes paired with microarchitectural implementations~\cite{lt-pim, bfree, pluto}, especially in binarized formats~\cite{biqgemm, subbitnn, tmac, luttensorcore}.
BiQGEMM \cite{biqgemm} divides the quantized weight matrix into sub-vectors and aggregates them to access the pre-computed lookup tables.
\cite{subbitnn} prepares several sub-bit kernels from all possible cases of binarized CNN filters.
T-MAC \cite{tmac} proposes a scalable mixed-bitwidth solution for arbitrary inputs.
To utilize LUTs for any pair of bitwidths, the weights are decomposed in a bit-wise manner.
%
LUT Tensor Core~\cite{luttensorcore} compresses the LUT by half using the symmetrical characteristic of integer representation.
While these approaches are well-suited for embedding LUTs on logic-based chips, they may not necessarily result in optimal performance on DRAM-PIM. 

\textbf{Low-bit quantization.}
Quantization is a well-known method that reduces data size. 
Reduced weights and activations reduce data transfer time and enable on-device storage \cite{quip2, shen2020q, binarybert, ternarybert, qimera, mimiq}.
BinaryBERT \cite{binarybert} quantizes the weights to 1-bit, achieving 24$\times$ memory reduction, while TernaryBERT~\cite{ternarybert} uses 3-bit weight quantization with knowledge distillation to maintain accuracy.
Furthermore, quantization lowers the computation burden.
I-BERT~\cite{ibert} employs integer-only quantization for hardware efficiency, and BiBERT~\cite{bibert} reduces the bitwidth to 1.
I-LLM~\cite{illm} quantizes the bitwidths of weights and activations to four with negligible information loss.
Similar trends can also be found on floating points, such as FP4~\cite{llmfp4} or hybrid formats~\cite{nvfp4, mx4, amxfp4}.
Since LUTs treat numbers as symbols, applying them to floating-point variants presents an opportunity for future work.

\section{Conclusion}

In this work, we present \thiswork, a novel LUT-based PIM architecture tailored for efficient low-bit quantized DNN inference. 
The key technique of \thiswork is to pack multiple operations in an LUT, to enable the inherent tradeoff between memory capacity and computational throughput in DRAM-PIM systems. 
{For a better tradeoff, we propose a set of techniques that reduce the growth rate of the LUT, and utilize the internal memory hierarchy of the DRAM-PIM.}
Our evaluation on real UPMEM PIM hardware demonstrates that \thiswork achieves significant performance improvements across a range of models and quantization settings. 

\section*{Acknowledgements}
This work was partially supported by 
National Research Foundation of Korea (NRF) grant funded by the Korea government (MSIT) (2022R1C1C1011307), 
and
Institute of Information \& communications Technology Planning \& Evaluation (IITP) (RS-2024-00395134, 
RS-2024-00347394, 
RS-2023-00256081, 
RS-2021-II211343).   
Jinho Lee is the corresponding author.



\bibliographystyle{IEEEtranS}
\bibliography{refs}

@INPROCEEDINGS {piccolo,
author = {Changmin Shin and Jaeyong Song and Hongsun Jang and Dogeun Kim and Jun Sung and Taehee Kwon and Jae Hyung Ju and Frank Liu and Yeonkyu Choi and Jinho Lee},
booktitle   = {HPCA},
title = {Piccolo: Large-Scale Graph Processing with Fine-Grained In-Memory Scatter-Gather},
year = {2025}
}

@inproceedings{fala,
author = {Shin, Changmin and Song, Jaeyong and Na, Seongmin and Sung, Jun and Jang, Hongsun and Lee, Jinho},
title = {FALA: Locality-Aware PIM-Host Cooperation for Graph Processing with Fine-Grained Column Access},
year = {2025},
booktitle = {MICRO}
}

@inproceedings{sub4_1,
 author = {Kim, Jeonghoon and Lee, Jung Hyun and Kim, Sungdong and Park, Joonsuk and Yoo, Kang Min and Kwon, Se Jung and Lee, Dongsoo},
 booktitle = {NeurIPS},
 title = {Memory-Efficient Fine-Tuning of Compressed Large Language Models via sub-4-bit Integer Quantization},
 year = {2023}
}

@INPROCEEDINGS{lowbitquant,
  author={Choukroun, Yoni and Kravchik, Eli and Yang, Fan and Kisilev, Pavel},
  booktitle={ICCVW}, 
  title={Low-bit Quantization of Neural Networks for Efficient Inference}, 
  year={2019},
  volume={},
  number={},
  pages={3009-3018},
  doi={10.1109/ICCVW.2019.00363}}

@inproceedings{2bitquant,
 author = {Choi, Jungwook and Venkataramani, Swagath and Srinivasan, Vijayalakshmi (Viji) and Gopalakrishnan, Kailash and Wang, Zhuo and Chuang, Pierce},
 booktitle = {MLSys},
 title = {Accurate and Efficient 2-bit Quantized Neural Networks},
 year = {2019}
}

@inproceedings{2bitquant_2,
 author = {Chee, Jerry and Cai, Yaohui and Kuleshov, Volodymyr and De Sa, Christopher M},
  booktitle={NeurIPS},
 editor = {A. Oh and T. Naumann and A. Globerson and K. Saenko and M. Hardt and S. Levine},
 title = {QuIP: 2-Bit Quantization of Large Language Models With Guarantees},
 year = {2023}
}

@INPROCEEDINGS{3bitquant,
  author={Nayak, Prateeth and Zhang, David and Chai, Sek},
  booktitle={EMC2-NIPS}, 
  title={Bit Efficient Quantization for Deep Neural Networks}, 
  year={2019},
  volume={},
  number={},
  pages={52-56},
  doi={10.1109/EMC2-NIPS53020.2019.00020}}

@inproceedings{4bitquant,
 author = {Xi, Haocheng and Li, ChangHao and Chen, Jianfei and Zhu, Jun},
 booktitle = {NeurIPS},
 editor = {A. Oh and T. Naumann and A. Globerson and K. Saenko and M. Hardt and S. Levine},
 pages = {49146--49168},
 publisher = {Curran Associates, Inc.},
 title = {Training Transformers with 4-bit Integers},
 year = {2023}
}

@inproceedings{sub4_2,
    title = "{B}it{D}istiller: Unleashing the Potential of Sub-4-Bit {LLM}s via Self-Distillation",
    author = "Du, DaYou  and
      Zhang, Yijia  and
      Cao, Shijie  and
      Guo, Jiaqi  and
      Cao, Ting  and
      Chu, Xiaowen  and
      Xu, Ningyi",
    booktitle = "ACL",
    year = "2024",
}

@INPROCEEDINGS{hbmpim,
  author={Kim, Jin Hyun and Kang, Shin-haeng and Lee, Sukhan and Kim, Hyeonsu and Song, Woongjae and Ro, Yuhwan and Lee, Seungwon and Wang, David and Shin, Hyunsung and Phuah, Bengseng and Choi, Jihyun and So, Jinin and Cho, YeonGon and Song, JoonHo and Choi, Jangseok and Cho, Jeonghyeon and Sohn, Kyomin and Sohn, Youngsoo and Park, Kwangil and Kim, Nam Sung},
  booktitle={Hot Chips}, 
  title={{Aquabolt-XL: Samsung HBM2-PIM with In-Memory Processing for ML Accelerators and Beyond}}, 
  year={2021}
}

@INPROCEEDINGS{aim,
  author={Kwon, Yongkee and Vladimir, Kornijcuk and Kim, Nahsung and Shin, Woojae and Won, Jongsoon and Lee, Minkyu and Joo, Hyunha and Choi, Haerang and Kim, Guhyun and An, Byeongju and Kim, Jeongbin and Lee, Jaewook and Kim, Ilkon and Park, Jaehan and Park, Chanwook and Song, Yosub and Yang, Byeongsu and Lee, Hyungdeok and Kim, Seho and Kwon, Daehan and Lee, Seongju and Kim, Kyuyoung and Oh, Sanghoon and Park, Joonhong and Hong, Gimoon and Ka, Dongyoon and Hwang, Kyudong and Park, Jeongje and Kang, Kyeongpil and Kim, Jungyeon and Jeon, Junyeol and Lee, Myeongjun and Shin, Minyoung and Shin, Minhwan and Cha, Jaekyung and Jung, Changson and Chang, Kijoon and Jeong, Chunseok and Lim, Euicheol and Park, Il and Chun, Junhyun and Hynix, Sk},
  booktitle={Hot Chips}, 
  title={{System Architecture and Software Stack for GDDR6-AiM}}, 
  year={2022}
}

@INPROCEEDINGS{chameleon,
  author={Asghari-Moghaddam, Hadi and Son, Young Hoon and Ahn, Jung Ho and Kim, Nam Sung},
  booktitle={MICRO}, 
  title={{Chameleon: Versatile and practical near-DRAM acceleration architecture for large memory systems}}, 
  year={2016}
}

@INPROCEEDINGS{tensordimm,
author = {Kwon, Youngeun and Lee, Yunjae and Rhu, Minsoo},
title = {{TensorDIMM: A Practical Near-Memory Processing Architecture for Embeddings and Tensor Operations in Deep Learning}},
year = {2019},
booktitle = {MICRO}
}

@INPROCEEDINGS{truepim,
  title={{The True Processing In Memory Accelerator}},
  author={Devaux, Fabrice},
  booktitle={Hot Chips},
  year={2019}
}

@INPROCEEDINGS{newton,
  author={He, Mingxuan and Song, Choungki and Kim, Ilkon and Jeong, Chunseok and Kim, Seho and Park, Il and Thottethodi, Mithuna and Vijaykumar, T. N.},
  booktitle={MICRO}, 
  title={{Newton: A DRAM-maker’s Accelerator-in-Memory (AiM) Architecture for Machine Learning}}, 
  year={2020}
}

@INPROCEEDINGS{gddraim,
  author={Lee, Seongju and Kim, Kyuyoung and Oh, Sanghoon and Park, Joonhong and Hong, Gimoon and Ka, Dongyoon and Hwang, Kyudong and Park, Jeongje and Kang, Kyeongpil and Kim, Jungyeon and Jeon, Junyeol and Kim, Nahsung and Kwon, Yongkee and Vladimir, Kornijcuk and Shin, Woojae and Won, Jongsoon and Lee, Minkyu and Joo, Hyunha and Choi, Haerang and Lee, Jaewook and Ko, Donguc and Jun, Younggun and Cho, Keewon and Kim, Ilwoong and Song, Choungki and Jeong, Chunseok and Kwon, Daehan and Jang, Jieun and Park, Il and Chun, Junhyun and Cho, Joohwan},
  booktitle={ISSCC}, 
  title={{A 1ynm 1.25V 8Gb, 16Gb/s/pin GDDR6-based Accelerator-in-Memory supporting 1TFLOPS MAC Operation and Various Activation Functions for Deep-Learning Applications}}, 
  year={2022}
}

@INPROCEEDINGS{ambit,
author = {Seshadri, Vivek and Lee, Donghyuk and Mullins, Thomas and Hassan, Hasan and Boroumand, Amirali and Kim, Jeremie and Kozuch, Michael A. and Mutlu, Onur and Gibbons, Phillip B. and Mowry, Todd C.},
title = {{Ambit: In-Memory Accelerator for Bulk Bitwise Operations using Commodity DRAM Technology}},
year = {2017},
booktitle = {MICRO}
}

@inproceedings{pidcomm,
  title={{PID-Comm: A Fast and Flexible Collective Communication Framework for Commodity Processing-in-DIMM Devices}},
  author={Noh, Si Ung and Hong, Junguk and Lim, Chaemin and Park, Seongyeon and Kim, Jeehyun and Kim, Hanjun and Kim, Youngsok and Lee, Jinho},
  booktitle={ISCA},
  year={2024}
}

@inproceedings{mimiq,
  title={MimiQ: Low-Bit Data-Free Quantization of Vision Transformers with Encouraging Inter-Head Attention Similarity},
  author={Choi, Kanghyun and Lee, Hyeyoon and Kwon, Dain and Park, SunJong and Kim, Kyuyeun and Park, Noseong and Choi, Jonghyun and Lee, Jinho},
  booktitle={Proceedings of the AAAI Conference on Artificial Intelligence},
  year={2025}
}

@inproceedings{pathweaver,
  title={$\{$PathWeaver$\}$: A $\{$High-Throughput$\}$$\{$Multi-GPU$\}$ System for $\{$Graph-Based$\}$ Approximate Nearest Neighbor Search},
  author={Kim, Sukjin and Park, Seongyeon and Noh, Si Ung and Hong, Junguk and Kwon, Taehee and Lim, Hunseong and Lee, Jinho},
  booktitle={2025 USENIX Annual Technical Conference (USENIX ATC 25)},
  year={2025}
}

@inproceedings{pimdl,
author = {Li, Cong and Zhou, Zhe and Wang, Yang and Yang, Fan and Cao, Ting and Yang, Mao and Liang, Yun and Sun, Guangyu},
title = {PIM-DL: Expanding the Applicability of Commodity DRAM-PIMs for Deep Learning via Algorithm-System Co-Optimization},
year = {2024},
booktitle={ASPLOS}
}

@inproceedings{pidjoin,
  title={Design and Analysis of a Processing-in-DIMM Join Algorithm: A Case Study with UPMEM DIMMs},
  author={Lim, Chaemin and Lee, Suhyun and Choi, Jinwoo and Lee, Jounghoo and Park, Seongyeon and Kim, Hanjun and Lee, Jinho and Kim, Youngsok},
  booktitle={SIGMOD},
  year={2023},
}

@inproceedings{spidjoin,
author = {Lee, Suhyun and Lim, Chaemin and Choi, Jinwoo and Choi, Heelim and Lee, Chan and Park, Yongjun and Park, Kwanghyun and Kim, Hanjun and Kim, Youngsok},
title = {SPID-Join: A Skew-resistant Processing-in-DIMM Join Algorithm Exploiting the Bank- and Rank-level Parallelisms of DIMMs},
year = {2024},
booktitle={SIGMOD}
}

@misc{pimdal,
      title={PIMDAL: Mitigating the Memory Bottleneck in Data Analytics using a Real Processing-in-Memory System}, 
      author={Manos Frouzakis and Juan Gómez-Luna and Geraldo F. Oliveira and Mohammad Sadrosadati and Onur Mutlu},
      year={2025},
      eprint={2504.01948},
      archivePrefix={arXiv},
      primaryClass={cs.AR},
      url={https://arxiv.org/abs/2504.01948}, 
}

@article{PyGim,
author = {Giannoula, Christina and Yang, Peiming and Fernandez, Ivan and Yang, Jiacheng and Durvasula, Sankeerth and Li, Yu Xin and Sadrosadati, Mohammad and Luna, Juan Gomez and Mutlu, Onur and Pekhimenko, Gennady},
title = {PyGim: An Efficient Graph Neural Network Library for Real Processing-In-Memory Architectures},
year = {2024},
issue_date = {December 2024},
publisher = {Association for Computing Machinery},
address = {New York, NY, USA},
volume = {8},
number = {3},
journal = {ACM POMACS},
month = dec,
articleno = {43},
numpages = {36}
}

@inproceedings {upmem_casestudy,
author = {Joel Nider and Craig Mustard and Andrada Zoltan and John Ramsden and Larry Liu and Jacob Grossbard and Mohammad Dashti and Romaric Jodin and Alexandre Ghiti and Jordi Chauzi and Alexandra Fedorova},
title = {A Case Study of {Processing-in-Memory} in {off-the-Shelf} Systems},
booktitle = {USENIX ATC},
year = {2021},
}

@INPROCEEDINGS{upimulator,
  author={Hyun, Bongjoon and Kim, Taehun and Lee, Dongjae and Rhu, Minsoo},
  booktitle={HPCA}, 
  title={Pathfinding Future PIM Architectures by Demystifying a Commercial PIM Technology}, 
  year={2024},
  volume={},
  number={},
  pages={263-279},
  doi={10.1109/HPCA57654.2024.00029}}

@INPROCEEDINGS{swiftrl,
  author={Gogineni, Kailash and Dayapule, Sai Santosh and Gómez-Luna, Juan and Gogineni, Karthikeya and Wei, Peng and Lan, Tian and Sadrosadati, Mohammad and Mutlu, Onur and Venkataramani, Guru},
  booktitle={ISPASS}, 
  title={SwiftRL: Towards Efficient Reinforcement Learning on Real Processing-In-Memory Systems}, 
  year={2024},
  volume={},
  number={},
  pages={217-229},
  doi={10.1109/ISPASS61541.2024.00029}}

@INPROCEEDINGS{upmem_benchmark,
  author={Gómez-Luna, Juan and El Hajj, Izzat and Fernandez, Ivan and Giannoula, Christina and Oliveira, Geraldo F. and Mutlu, Onur},
  booktitle={IGSC}, 
  title={Benchmarking Memory-Centric Computing Systems: Analysis of Real Processing-In-Memory Hardware},
  year={2021}
}

@INPROCEEDINGS{simplepim,
  author={Chen, Jinfan and Gómez-Luna, Juan and El Hajj, Izzat and Guo, Yuxin and Mutlu, Onur},
  booktitle={PACT}, 
  title={SimplePIM: A Software Framework for Productive and Efficient Processing-in-Memory}, 
  year={2023}
}

@INPROCEEDINGS{virtualpim,
  author={Kim, Donghyeon and Kim, Taehoon and Hwang, Inyong and Park, Taehyeong and Kim, Hanjun and Kim, Youngsok and Park, Yongjun},
  booktitle={PACT}, 
  title={Virtual PIM: Resource-Aware Dynamic DPU Allocation and Workload Scheduling Framework for Multi-DPU PIM Architecture}, 
  year={2023},}

@INPROCEEDINGS{transpimlib,
  author={Item, Maurus and Oliveira, Geraldo F. and Gómez-Luna, Juan and Sadrosadati, Mohammad and Guo, Yuxin and Mutlu, Onur},
  booktitle={ISPASS}, 
  title={TransPimLib: Efficient Transcendental Functions for Processing-in-Memory Systems}, 
  year={2023},
  volume={},
  number={},
  pages={235-247},
  doi={10.1109/ISPASS57527.2023.00031}}

@ARTICLE{axdimm,
  author={Ke, Liu and Zhang, Xuan and So, Jinin and Lee, Jong-Geon and Kang, Shin-Haeng and Lee, Sukhan and Han, Songyi and Cho, YeonGon and Kim, Jin Hyun and Kwon, Yongsuk and Kim, KyungSoo and Jung, Jin and Yun, Ilkwon and Park, Sung Joo and Park, Hyunsun and Song, Joonho and Cho, Jeonghyeon and Sohn, Kyomin and Kim, Nam Sung and Lee, Hsien-Hsin S.},
  journal={IEEE Micro}, 
  title={Near-Memory Processing in Action: Accelerating Personalized Recommendation With AxDIMM}, 
  year={2022}
}

@ARTICLE{lazypim,
  author={Boroumand, Amirali and Ghose, Saugata and Patel, Minesh and Hassan, Hasan and Lucia, Brandon and Hsieh, Kevin and Malladi, Krishna T. and Zheng, Hongzhong and Mutlu, Onur},
  journal={IEEE CAL}, 
  title={{LazyPIM: An Efficient Cache Coherence Mechanism for Processing-in-Memory}}, 
  year={2017}
}

@article{cacti,
author = {Balasubramonian, Rajeev and Kahng, Andrew B. and Muralimanohar, Naveen and Shafiee, Ali and Srinivas, Vaishnav},
title = {{CACTI 7: New Tools for Interconnect Exploration in Innovative Off-Chip Memories}},
year = {2017},
journal = {ACM TACO}
}

@INPROCEEDINGS{gradpim,
  author={Kim, Heesu and Park, Hanmin and Kim, Taehyun and Cho, Kwanheum and Lee, Eojin and Ryu, Soojung and Lee, Hyuk-Jae and Choi, Kiyoung and Lee, Jinho},
  booktitle={HPCA}, 
  title={{GradPIM: A Practical Processing-in-DRAM Architecture for Gradient Descent}}, 
  year={2021}
}

@inproceedings{bufcmp,
  title={Buffered compares: Excavating the hidden parallelism inside DRAM architectures with lightweight logic},
  author={Lee, Jinho and Ahn, Jung Ho and Choi, Kiyoung},
  booktitle={DATE},
  year={2016}
}

@inproceedings{trim,
author = {Park, Jaehyun and Kim, Byeongho and Yun, Sungmin and Lee, Eojin and Rhu, Minsoo and Ahn, Jung Ho},
title = {{TRiM: Enhancing Processor-Memory Interfaces with Scalable Tensor Reduction in Memory}},
year = {2021},
booktitle = {MICRO}
}

@article{mvid,
  title={MViD: Sparse matrix-vector multiplication in mobile dram for accelerating recurrent neural networks},
  author={Kim, Byeongho and Chung, Jongwook and Lee, Eojin and Jung, Wonkyung and Lee, Sunjung and Choi, Jaewan and Park, Jaehyun and Wi, Minbok and Lee, Sukhan and Ahn, Jung Ho},
  journal={IEEE TC},
  volume={69},
  number={7},
  pages={955--967},
  year={2020},
  publisher={IEEE}
}

@inproceedings{attacc,
author = {Park, Jaehyun and Choi, Jaewan and Kyung, Kwanhee and Kim, Michael Jaemin and Kwon, Yongsuk and Kim, Nam Sung and Ahn, Jung Ho},
title = {AttAcc! Unleashing the Power of PIM for Batched Transformer-based Generative Model Inference},
year = {2024},
booktitle = {ASPLOS}
}

@inproceedings{duplex,
title={Duplex: A Device for Large Language Models with Mixture of Experts, Grouped Query Attention, and Continuous Batching}, 
author={Sungmin Yun and Kwanhee Kyung and Juhwan Cho and Jaewan Choi and Jongmin Kim and Byeongho Kim and Sukhan Lee and Kyomin Sohn and Jung Ho Ahn},
year={2024},
booktitle = {MICRO}
}

@inproceedings{aespa,
author = {Kal, Hongju and Yoo, Chanyoung and Ro, Won Woo},
title = {AESPA: Asynchronous Execution Scheme to Exploit Bank-Level Parallelism of Processing-in-Memory},
year = {2023},
booktitle = {MICRO}
}

@INPROCEEDINGS{psyncpim,
  author={Baek, Daehyeon and Hwang, Soojin and Huh, Jaehyuk},
  booktitle={ISCA}, 
  title={pSyncPIM: Partially Synchronous Execution of Sparse Matrix Operations for All-Bank PIM Architectures}, 
  year={2024}
}

@inproceedings{neupims,
author = {Heo, Guseul and Lee, Sangyeop and Cho, Jaehong and Choi, Hyunmin and Lee, Sanghyeon and Ham, Hyungkyu and Kim, Gwangsun and Mahajan, Divya and Park, Jongse},
title = {NeuPIMs: NPU-PIM Heterogeneous Acceleration for Batched LLM Inferencing},
year = {2024},
booktitle = {ASPLOS}
}

@inproceedings{ianus,
author = {Seo, Minseok and Nguyen, Xuan Truong and Hwang, Seok Joong and Kwon, Yongkee and Kim, Guhyun and Park, Chanwook and Kim, Ilkon and Park, Jaehan and Kim, Jeongbin and Shin, Woojae and Won, Jongsoon and Choi, Haerang and Kim, Kyuyoung and Kwon, Daehan and Jeong, Chunseok and Lee, Sangheon and Choi, Yongseok and Byun, Wooseok and Baek, Seungcheol and Lee, Hyuk-Jae and Kim, John},
title = {IANUS: Integrated Accelerator based on NPU-PIM Unified Memory System},
year = {2024},
booktitle = {ASPLOS}
}

@inproceedings{samsung-pim,
  title={Hardware architecture and software stack for PIM based on commercial DRAM technology: Industrial product},
  author={Lee, Sukhan and Kang, Shin-haeng and Lee, Jaehoon and Kim, Hyeonsu and Lee, Eojin and Seo, Seungwoo and Yoon, Hosang and Lee, Seungwon and Lim, Kyounghwan and Shin, Hyunsung and others},
  booktitle={ISCA},
  year={2021}
}

@INPROCEEDINGS{hbmpim_isscc,
  author={Kwon, Young-Cheon and Lee, Suk Han and Lee, Jaehoon and Kwon, Sang-Hyuk and Ryu, Je Min and Son, Jong-Pil and Seongil, O and Yu, Hak-Soo and Lee, Haesuk and Kim, Soo Young and Cho, Youngmin and Kim, Jin Guk and Choi, Jongyoon and Shin, Hyun-Sung and Kim, Jin and Phuah, BengSeng and Kim, HyoungMin and Song, Myeong Jun and Choi, Ahn and Kim, Daeho and Kim, SooYoung and Kim, Eun-Bong and Wang, David and Kang, Shinhaeng and Ro, Yuhwan and Seo, Seungwoo and Song, JoonHo and Youn, Jaeyoun and Sohn, Kyomin and Kim, Nam Sung},
  booktitle={ISSCC}, 
  title={25.4 A 20nm 6GB Function-In-Memory DRAM, Based on HBM2 with a 1.2TFLOPS Programmable Computing Unit Using Bank-Level Parallelism, for Machine Learning Applications}, 
  year={2021}
}

@inproceedings{lutnn,
author = {Tang, Xiaohu and Wang, Yang and Cao, Ting and Zhang, Li Lyna and Chen, Qi and Cai, Deng and Liu, Yunxin and Yang, Mao},
title = {LUT-NN: Empower Efficient Neural Network Inference with Centroid Learning and Table Lookup},
year = {2023},
booktitle = {MobiCom}
}

@INPROCEEDINGS{rowclone,
  author={Seshadri, Vivek and Kim, Yoongu and Fallin, Chris and Lee, Donghyuk and Ausavarungnirun, Rachata and Pekhimenko, Gennady and Luo, Yixin and Mutlu, Onur and Gibbons, Phillip B. and Kozuch, Michael A. and Mowry, Todd C.},
  booktitle={MICRO}, 
  title={RowClone: Fast and energy-efficient in-DRAM bulk data copy and initialization}, 
  year={2013}
}

@INPROCEEDINGS{asyncdimm,
  author={Chen, Liyan and Lyu, Dongxu and Jiang, Jianfei and Wang, Qin and Mao, Zhigang and Jing, Naifeng},
  booktitle={HPCA}, 
  title={AsyncDIMM: Achieving Asynchronous Execution in DIMM-Based Near-Memory Processing}, 
  year={2025}
}

@inproceedings{simdram,
author = {Hajinazar, Nastaran and Oliveira, Geraldo F. and Gregorio, Sven and Ferreira, Jo\~{a}o Dinis and Ghiasi, Nika Mansouri and Patel, Minesh and Alser, Mohammed and Ghose, Saugata and G\'{o}mez-Luna, Juan and Mutlu, Onur},
title = {SIMDRAM: a framework for bit-serial SIMD processing using DRAM},
year = {2021},
booktitle = {ASPLOS},
}

@INPROCEEDINGS{mimdram,
  author={Oliveira, Geraldo F. and Olgun, Ataberk and Yağlıkçı, Abdullah Giray and Bostancı, F. Nisa and Gómez-Luna, Juan and Ghose, Saugata and Mutlu, Onur},
  booktitle={HPCA}, 
  title={MIMDRAM: An End-to-End Processing-Using-DRAM System for High-Throughput, Energy-Efficient and Programmer-Transparent Multiple-Instruction Multiple-Data Computing}, 
  year={2024}
}

@INPROCEEDINGS{lutdla,
  author={Li, Guoyu and Ye, Shengyu and Chen, Chunyun and Wang, Yang and Yang, Fan and Cao, Ting and Liu, Cheng and Aly, Mohamed M. Sabry and Yang, Mao},
  booktitle={HPCA}, 
  title={LUT-DLA: Lookup Table as Efficient Extreme Low-Bit Deep Learning Accelerator}, 
  year={2025}
}

@inproceedings{luttensorcore,
author = {Mo, Zhiwen and Wang, Lei and Wei, Jianyu and Zeng, Zhichen and Cao, Shijie and Ma, Lingxiao and Jing, Naifeng and Cao, Ting and Xue, Jilong and Yang, Fan and Yang, Mao},
title = {LUT Tensor Core: A Software-Hardware Co-Design for LUT-Based Low-Bit LLM Inference},
booktitle={ISCA},
year = {2025}
}

@INPROCEEDINGS{pnmpudpim,
  author={Mutlu, Onur and Olgun, Ataberk and Y{\"u}ksel, \.{I}smail Emir},
  booktitle={IMW}, 
  title={Memory-Centric Computing: Solving Computing’s Memory Problem}, 
  year={2025}
}

@inproceedings{tmac,
author = {Wei, Jianyu and Cao, Shijie and Cao, Ting and Ma, Lingxiao and Wang, Lei and Zhang, Yanyong and Yang, Mao},
title = {T-MAC: CPU Renaissance via Table Lookup for Low-Bit LLM Deployment on Edge},
year = {2025},
booktitle={EuroSys}
}

@InProceedings{subbitnn,
    author    = {Wang, Yikai and Yang, Yi and Sun, Fuchun and Yao, Anbang},
    title     = {Sub-Bit Neural Networks: Learning To Compress and Accelerate Binary Neural Networks},
    booktitle = {ICCV},
    month     = {October},
    year      = {2021},
    pages     = {5360-5369}
}

@INPROCEEDINGS{biqgemm,
  author={Jeon, Yongkweon and Park, Baeseong and Kwon, Se Jung and Kim, Byeongwook and Yun, Jeongin and Lee, Dongsoo},
  booktitle={SC}, 
  title={BiQGEMM: Matrix Multiplication with Lookup Table for Binary-Coding-Based Quantized DNNs}, 
  year={2020},
  volume={},
  number={},
  pages={1-14},
  doi={10.1109/SC41405.2020.00099}}

@inproceedings{ibert,
  title={I-bert: Integer-only bert quantization},
  author={Kim, Sehoon and Gholami, Amir and Yao, Zhewei and Mahoney, Michael W and Keutzer, Kurt},
  booktitle={ICML},
  year={2021},
  organization={PMLR}
}

@article{qimera,
  title={{Qimera: Data-free quantization with synthetic boundary supporting samples}},
  author={Choi, Kanghyun and Hong, Deokki and Park, Noseong and Kim, Youngsok and Lee, Jinho},
  journal={NeurIPS},
  year={2021}
}

@article{bibert,
  title={Bibert: Accurate fully binarized bert},
  author={Qin, Haotong and Ding, Yifu and Zhang, Mingyuan and Yan, Qinghua and Liu, Aishan and Dang, Qingqing and Liu, Ziwei and Liu, Xianglong},
  journal={arXiv preprint arXiv:2203.06390},
  year={2022}
}

@article{illm,
  title={I-llm: Efficient integer-only inference for fully-quantized low-bit large language models},
  author={Hu, Xing and Cheng, Yuan and Yang, Dawei and Yuan, Zhihang and Yu, Jiangyong and Xu, Chen and Zhou, Sifan},
  journal={arXiv preprint arXiv:2405.17849},
  year={2024}
}

@inproceedings{shen2020q,
  title={Q-bert: Hessian based ultra low precision quantization of bert},
  author={Shen, Sheng and Dong, Zhen and Ye, Jiayu and Ma, Linjian and Yao, Zhewei and Gholami, Amir and Mahoney, Michael W and Keutzer, Kurt},
  booktitle={AAAI},

  year={2020}
}

@article{kdlsq,
  title={Kdlsq-bert: A quantized bert combining knowledge distillation with learned step size quantization},
  author={Jin, Jing and Liang, Cai and Wu, Tiancheng and Zou, Liqin and Gan, Zhiliang},
  journal={arXiv preprint arXiv:2101.05938},
  year={2021}
}

@article{ternarybert,
  title={Ternarybert: Distillation-aware ultra-low bit bert},
  author={Zhang, Wei and Hou, Lu and Yin, Yichun and Shang, Lifeng and Chen, Xiao and Jiang, Xin and Liu, Qun},
  journal={arXiv preprint arXiv:2009.12812},
  year={2020}
}

@inproceedings{quip2,
  title={Quip: 2-bit quantization of large language models with guarantees},
  author={Chee, Jerry and Cai, Yaohui and Kuleshov, Volodymyr and De Sa, Christopher M},
  booktitle={NeurIPS},
  year={2023}
}

@article{binarybert,
  title={Binarybert: Pushing the limit of bert quantization},
  author={Bai, Haoli and Zhang, Wei and Hou, Lu and Shang, Lifeng and Jin, Jing and Jiang, Xin and Liu, Qun and Lyu, Michael and King, Irwin},
  journal={arXiv preprint arXiv:2012.15701},
  year={2020}
}

@ARTICLE{hbm3_jssc,
  author={Park, Myeong-Jae and Lee, Jinhyung and Cho, Kyungjun and Park, Jihwan and Moon, Junil and Lee, Sung-Hak and Kim, Tae-Kyun and Oh, Sanghoon and Choi, Seokwoo and Choi, Yongsuk and Cho, Ho Sung and Yun, Taesik and Koo, Young Jun and Lee, Jae-Seung and Yoon, Byung-Kuk and Park, Young-Jun and Oh, Sangmuk and Lee, Chang Kwon and Lee, Seong-Hee and Kim, Hyun-Woo and Ju, Yucheon and Lim, Seung-Kyun and Lee, Kyo Yun and Lee, Sang-Hoon and We, Woo Sung and Kim, Seungchan and Yang, Seung Min and Lee, Keonho and Kim, In-Keun and Jeon, Younghyun and Park, Jae-Hyung and Yun, Jong Chan and Kim, Seonyeol and Lee, Dong-Yeol and Oh, Su-Hyun and Shin, Jung-Hyun and Lee, Yeonho and Jang, Jieun and Cho, Joohwan},
  journal={IEEE JSSC}, 
  title={A 192-Gb 12-High 896-GB/s HBM3 DRAM With a TSV Auto-Calibration Scheme and Machine-Learning-Based Layout Optimization}, 
  year={2023},
  volume={58},
  number={1}
}

@ARTICLE{ramulator2,
  author={Luo, Haocong and Tuğrul, Yahya Can and Bostancı, F. Nisa and Olgun, Ataberk and Yağlıkçı, A. Giray and Mutlu, Onur},
  journal={IEEE CAL}, 
  title={Ramulator 2.0: A Modern, Modular, and Extensible DRAM Simulator}, 
  year={2024},
  volume={23},
  number={1}
}

@article{lt-pim,
  title={Lt-pim: An lut-based processing-in-dram architecture with rowhammer self-tracking},
  author={Zhou, Ranyang and Tabrizchi, Sepehr and Roohi, Arman and Angizi, Shaahin},
  journal={IEEE CAL},
  volume={21},
  number={2},
  year={2022}
}

@inproceedings{bfree,
  title={Look-up table based energy efficient processing in cache support for neural network acceleration},
  author={Ramanathan, Akshay Krishna and Kalsi, Gurpreet S and Srinivasa, Srivatsa and Chandran, Tarun Makesh and Pillai, Kamlesh R and Omer, Om J and Narayanan, Vijaykrishnan and Subramoney, Sreenivas},
  booktitle={MICRO},
  year={2020},
}

@INPROCEEDINGS{pluto,
  author={Ferreira, João Dinis and Falcao, Gabriel and Gómez-Luna, Juan and Alser, Mohammed and Orosa, Lois and Sadrosadati, Mohammad and Kim, Jeremie S. and Oliveira, Geraldo F. and Shahroodi, Taha and Nori, Anant and Mutlu, Onur},
  booktitle={2022 55th IEEE/ACM International Symposium on Microarchitecture (MICRO)}, 
  title={pLUTo: Enabling Massively Parallel Computation in DRAM via Lookup Tables}, 
  year={2022},
  volume={},
  number={},
  pages={900-919},
  doi={10.1109/MICRO56248.2022.00067}}

@inproceedings{llmfp4,
  title={Llm-fp4: 4-bit floating-point quantized transformers},
  author={Liu, Shih-yang and Liu, Zechun and Huang, Xijie and Dong, Pingcheng and Cheng, Kwang-Ting},
  booktitle={EMNLP},
  year={2023}
}

@article{nvfp4,
  title={FP4 All the Way: Fully Quantized Training of LLMs},
  author={Chmiel, Brian and Fishman, Maxim and Banner, Ron and Soudry, Daniel},
  journal={arXiv preprint arXiv:2505.19115},
  year={2025}
}

@inproceedings{amxfp4,
  title={Amxfp4: Taming activation outliers with asymmetric microscaling floating-point for 4-bit llm inference},
  author={Lee, Janghwan and Park, Jiwoong and Kim, Jinseok and Kim, Yongjik and Oh, Jungju and Oh, Jinwook and Choi, Jungwook},
  booktitle={ACL findings},
  year={2025}
}

@article{mx4,
  title={Microscaling data formats for deep learning},
  author={Rouhani, Bita Darvish and Zhao, Ritchie and More, Ankit and Hall, Mathew and Khodamoradi, Alireza and Deng, Summer and Choudhary, Dhruv and Cornea, Marius and Dellinger, Eric and Denolf, Kristof and others},
  journal={arXiv preprint arXiv:2310.10537},
  year={2023}
}

@misc{nxfp,
    title={Nanoscaling Floating-Point (NxFP): NanoMantissa, Adaptive Microexponents, and Code Recycling for Direct-Cast Compression of Large Language Models},
    author={Yun-Chen Lo and Gu-Yeon Wei and David Brooks},
    year={2024},
    eprint={2412.19821},
    archivePrefix={arXiv},
    primaryClass={cs.AR}
}

@manual{upmem_sdk,
  title        = {UPMEM Software Development Kit (SDK)},
  author       = {{UPMEM}},
  year         = {2025},
  note         = {[Online; accessed 2025-08-01]},
  url          = {https://sdk.upmem.com/}
}

@inproceedings{smartinfinity,
  title={{Smart-Infinity: Fast Large Language Model Training using Near-Storage Processing on a Real System}},
  author={Jang, Hongsun and Song, Jaeyong and Jung, Jaewon and Park, Jaeyoung and Kim, Youngsok and Lee, Jinho},
  booktitle={HPCA},
  year={2024}
}

@inproceedings{blink,
  title={Blink: Fast and generic collectives for distributed ml},
  author={Wang, Guanhua and Venkataraman, Shivaram and Phanishayee, Amar and Devanur, Nikhil and Thelin, Jorgen and Stoica, Ion},
  booktitle={MLSys},
  year={2020}
}

@article{megatron-lm,
  title={{Megatron-lm: Training Multi-billion Parameter Language Models Using Model Parallelism}},
  author={Shoeybi, Mohammad and Patwary, Mostofa and Puri, Raul and LeGresley, Patrick and Casper, Jared and Catanzaro, Bryan},
  journal={arXiv preprint arXiv:1909.08053},
  year={2019}
}

@inproceedings{optimus-cc,
author = {Song, Jaeyong and Yim, Jinkyu and Jung, Jaewon and Jang, Hongsun and Kim, Hyung-Jin and Kim, Youngsok and Lee, Jinho},
title = {{Optimus-CC: Efficient Large NLP Model Training with 3D Parallelism Aware Communication Compression}},
year = {2023},
booktitle = {ASPLOS},
}

@inproceedings{sancus,
author = {Peng, Jingshu and Chen, Zhao and Shao, Yingxia and Shen, Yanyan and Chen, Lei and Cao, Jiannong},
title = {{Sancus: Staleness-Aware Communication-Avoiding Full-Graph Decentralized Training in Large-Scale Graph Neural Networks}},
year = {2022},
booktitle = {pVLDB}
}

@inproceedings{themis,
  title={Themis: A network bandwidth-aware collective scheduling policy for distributed training of dl models},
  author={Rashidi, Saeed and Won, William and Srinivasan, Sudarshan and Sridharan, Srinivas and Krishna, Tushar},
  booktitle={ISCA},
  year={2022}
}

@inproceedings{dimmlink,
  title={DIMM-Link: Enabling Efficient Inter-DIMM Communication for Near-Memory Processing},
  author={Zhou, Zhe and Li, Cong and Yang, Fan and Suny, Guangyu},
  booktitle={HPCA},
  year={2023},
}

@inproceedings{ndpbridge,
  title={Ndpbridge: Enabling cross-bank coordination in near-dram-bank processing architectures},
  author={Tian, Boyu and Li, Yiwei and Jiang, Li and Cai, Shuangyu and Gao, Mingyu},
  booktitle={ISCA},
  year={2024},
}

@inproceedings{flexram,
   Author = {Kang, Yi and Huang, Wei and Yoo, Seung-Moon and Keen, D and Ge, Zhenzhou and Lam, V and Pattnaik, P and Torrellas, J},
   Title = {{FlexRAM: toward an advanced intelligent memory system}},
   BookTitle = {ICCD},
      Year = {1999} }

@inproceedings{yukon,
   Author = {Kirsch, Graham},
   Title = {{Active Memory: Micron' s Yukon}},
   BookTitle = {IPDPS},
      Year = {2003} }

@inproceedings{execube,
   Author = {Kogge, Peter M},
   Title = {{EXECUBE-A new architecture for scaleable MPPs}},
   BookTitle = {ICPP},
      Year = {1994} }

@inproceedings{smartmem,
   Author = {Mai, Ken and Paaske, T and Jayasena, N and Ho, R and Dally, WJ and Horowitz, M},
   Title = {{Smart Memories: a modular reconfigurable architecture}},
   BookTitle = {ISCA},
      Year = {2000} }

@article{IRAM,
   Author = {Patterson, David and Anderson, Thomas and Cardwell, Neal and Fromm, Richard and Keeton, Kimberly and Kozyrakis, Christoforos and Thomas, Randi and Yelick, Katherine},
   Title = {{A case for intelligent RAM}},
   Journal = {IEEE Micro},
      Year = {1997} }

@article{transformer,
  title={Attention is all you need},
  author={Vaswani, Ashish and Shazeer, Noam and Parmar, Niki and Uszkoreit, Jakob and Jones, Llion and Gomez, Aidan N and Kaiser, {\L}ukasz and Polosukhin, Illia},
  journal={NeurIPS},
  year={2017}
}

@inproceedings{bert,
  title={Bert: Pre-training of deep bidirectional transformers for language understanding},
  author={Devlin, Jacob and Chang, Ming-Wei and Lee, Kenton and Toutanova, Kristina},
  booktitle={NAACL},
  year={2019}
}

@inproceedings{vit,
  title={An image is worth 16x16 words: Transformers for image recognition at scale},
  author={Dosovitskiy, Alexey and Beyer, Lucas and Kolesnikov, Alexander and Weissenborn, Dirk and Zhai, Xiaohua and Unterthiner, Thomas and Dehghani, Mostafa and Minderer, Matthias and Heigold, Georg and Gelly, Sylvain and others},
  booktitle={ICLR},
  year={2021}
}

@article{opt,
  title={Opt: Open pre-trained transformer language models},
  author={Zhang, Susan and Roller, Stephen and Goyal, Naman and Artetxe, Mikel and Chen, Moya and Chen, Shuohui and Dewan, Christopher and Diab, Mona and Li, Xian and Lin, Xi Victoria and others},
  journal={arXiv preprint arXiv:2205.01068},
  year={2022}
}

@inproceedings{imagenet,
  title={Imagenet: A large-scale hierarchical image database},
  author={Deng, Jia and Dong, Wei and Socher, Richard and Li, Li-Jia and Li, Kai and Fei-Fei, Li},
  booktitle={2009 IEEE conference on computer vision and pattern recognition},
  pages={248--255},
  year={2009},
  organization={Ieee}
}

@inproceedings{qvit,
  title={Q-vit: Accurate and fully quantized low-bit vision transformer},
  author={Li, Yanjing and Xu, Sheng and Zhang, Baochang and Cao, Xianbin and Gao, Peng and Guo, Guodong},
  booktitle={NeurIPS},
  year={2022}
}

@article{omniquant,
  title={Omniquant: Omnidirectionally calibrated quantization for large language models},
  author={Shao, Wenqi and Chen, Mengzhao and Zhang, Zhaoyang and Xu, Peng and Zhao, Lirui and Li, Zhiqian and Zhang, Kaipeng and Gao, Peng and Qiao, Yu and Luo, Ping},
  journal={arXiv preprint arXiv:2308.13137},
  year={2023}
}

@article{gunter2024apple,
  title={Apple intelligence foundation language models},
  author={Gunter, Tom and Wang, Zirui and Wang, Chong and Pang, Ruoming and Narayanan, Andy and Zhang, Aonan and Zhang, Bowen and Chen, Chen and Chiu, Chung-Cheng and Qiu, David and others},
  journal={arXiv preprint arXiv:2407.21075},
  year={2024}
}

@article{bitnet,
  title={1-bit ai infra: Part 1.1, fast and lossless bitnet b1. 58 inference on cpus},
  author={Wang, Jinheng and Zhou, Hansong and Song, Ting and Mao, Shaoguang and Ma, Shuming and Wang, Hongyu and Xia, Yan and Wei, Furu},
  journal={arXiv preprint arXiv:2410.16144},
  year={2024}
}

\clearpage
%
%
%
%
%



\appendix
\section{Artifact Appendix}

\subsection{Abstract}

{\em
We provide the source code of \thiswork along with additional scripts required for execution.
The artifact includes the procedures for generating both the \lut and the \reorder, as well as how these LUTs are utilized during LUT-based inference.
It also contains the \dataflow implemented in the PIM-side code.
Moreover, the source code includes baseline methods (e.g., naive MAC operations on PIM and activation-driven LUT approaches such as LUT Tensor Core) and the operation-packed LUT.
For the most recent version of \thiswork, please refer to the up-to-date artifact link.
}

\subsection{Artifact check-list (meta-information)}
{\small
\begin{itemize}
  \item {\bf Algorithm: }LUT-based GEMM operation.
  \item {\bf Run-time environment: }Ubuntu 18.04.
  \item {\bf Hardware: }DDR4 channels equipped with UPMEM DIMMs, multicore x86\_64 CPU.
  \item {\bf Metrics: }Execution time with functinoality check.
  \item {\bf How much disk space required (approximately)?: }1GB.
  \item {\bf How much time is needed to prepare workflow (approximately)?: }Several hours.
  \item {\bf How much time is needed to complete experiments (approximately)?: }It takes less than a minute.
  \item {\bf Publicly available?: }Yes.
  \item {\bf Archived DOI: }https://doi.org/10.5281/zenodo.17851318
\end{itemize}
}

\subsection{Description}

\subsubsection{How to access}

{
Please access the artifact through the archived DOI or our GitHub link.}

\subsubsection{Hardware dependencies}

{
DDR4 channels equipped with UPMEM DIMMs with multicore x86\_64 CPU.
Our evaluation was conducted using eight DDR4 channels equipped with UPMEM DIMMs and an Intel Xeon Gold 5215 CPU.
}

\subsubsection{Software dependencies}

{
The required software to run the source code of \thiswork is listed below.
{\small
\begin{itemize}
  \item UPMEM SDK driver (version 2023.2.0)
  \item g++
  \item Python 3.6
  \item Scipy
  \item Pandas
  \item Numpy
  \item Matplotlib
\end{itemize}
}
}

\subsubsection{Datasets}
{
Generated datasets were used to demonstrate the effectiveness of \thiswork.
Although, \thiswork focuses on low-bit quantization, the values of the generated elements remain within the representable rage defined by the activation and weight bitwidths.
Other datasets can also be used without losing generality.
}

\subsection{Installation}

{
The UPMEM SDK driver (version 2023.2.0) can be obtained from https://sdk.upmem.com/.
Once installed, the UPMEM\_HOME environment variable should be configured to point to the directory containing the library by sourcing the upmem\_env.sh script provided in the UPMEM-2023.2.0.
The artifact also operates correctly with other versions of the UPMEM SDK driver.
}

\subsection{Experiment workflow}

{
Although, \thiswork is primarily designed for GEMM operations, the provided code enables comparison against baselines, including the conventional MAC operations without the use of LUTs, and activation-driven LUT methods such as LUT Tensor Core.
We also compare \thiswork with an operation-packed LUT and a version of \thiswork without	\dataflow.
Using the {\em script.h} script, users can source the UPMEM driver, compile, and run LUT-based GEMM operations while varying multiple parameters.
This allows evaluation of \thiswork across different matrix dimensions (e.g., varying N/M/K).

To further explore the benefits and tradeoffs of LUT use, users can examine the impact of different packing degrees by adjusting the parameter PD in the script.
Additionally, the code includes procedures for generating both the \lut and the \reorder, considering the bitwidths of activations and weights.
During execution, the system allocates PEs, constructs the LUTs, transfers data to the PIM, performs the GEMM operation within the PIM through dpu binary codes, and returns the output for additional correctness verification.
}

\subsection{Evaluation and expected results}

{
The code will execute GEMM operations using various LUT-based methods or without LUT, demonstrating the efficiency of \thiswork through execution time and functionality checks.
}

\subsection{Notes}
{
For a detailed explanation of the procedures, please refer to the guidelines in the artifact's README file.
}



\end{document}